\documentclass[notoc,paper,letterpaper]{JHEP3}
\usepackage{amssymb}
\def\be{\begin{equation}}
\def\ee{\end{equation}}
\def\bea{\begin{eqnarray}}
\def\eea{\end{eqnarray}}
\def\ie{{\it i.e.}}
\def\eg{{\it e.g.}}
\def\etc{{\it etc.}}
\def\C{\mathbb{C}}
\def\CP{\mathbb{CP}}

\def\Z{\mathbb{Z}}
\def\sm{\small}
\def\U{\mbox{U}}
\def\SU{\mathop{\rm SU}}
\def\SO{\mathop{\rm SO}}
\def\SL{\mathop{\rm SL}}
\def\GL{\mathop{\rm GL}}
\def\Sp{\mathop{\rm Sp}}
\def\bo{\hbox{1\kern-.23em{\rm l}}}
\def\til{\widetilde}

\def\bar{\overline}
\def\del{{\partial}}

\def\Re{{\mbox{Re}}}
\def\Im{{\mbox{Im}}}
\def\a{{\alpha}}
\def\b{{\beta}}
\def\g{{\gamma}}
\def\d{{\delta}}

\def\l{{\lambda}}
\def\t{{\tau}}

\def\x{{\xi}}
\def\o{{\omega}}
\def\AA{{\cal A}}
\def\BB{{\cal B}}
\def\CC{{\cal C}}
\def\DD{{\cal D}}
\def\MM{{\cal M}}
\def\OO{{\cal O}}
\def\ib{{\bar\imath}}
\def\jb{{\bar\jmath}}
\def\ud{{\dot u}}
\def\udd{{\ddot u}}

\title{Classification of N=2 Superconformal Field Theories 
with Two-Dimensional Coulomb Branches}

\author{Philip C. Argyres$^{1,2}$, Michael Crescimanno$^3$, 
Alfred D. Shapere$^4$ and John R. Wittig$^1$\\ 
$^1$\,Physics Dept., Univ.\ of Cincinnati, 
	Cincinnati OH 45221-0011\\
\ \ \email{argyres,jwittig(at)physics.uc.edu}
\vskip5pt
$^2$\,School of Natural Sciences, Institute for Advanced Study, 
	Princeton NJ 08540
\vskip5pt
$^3$\,Dept.\ of Physics and Astronomy, Youngstown State Univ., 
	Youngstown OH 44555-2001\\
\ \ \email{mcrescim(at)as.ysu.edu}
\vskip5pt
$^4$\,Dept.\ of Physics and Astronomy, Univ.\ of Kentucky, 
	Lexington KY 40506-0055\\
\ \ \email{shapere(at)pa.uky.edu}}

\abstract{We study the classification of 2-dimensional scale-invariant 
rigid special Kahler (RSK) geometries, which potentially describe the 
Coulomb branches of $N{=}2$ supersymmetric field theories in four 
dimensions.  We show that this classification is equivalent to the 
solution of a set of polynomial equations by using an integrability 
condition for the 
central charge, scale invariance, constraints coming from demanding 
single-valuedness of physical quantities on the Coulomb branch, and 
properties of massless BPS states at singularities.  We find solutions 
corresponding 
to lagrangian scale invariant theories---including the scale invariant 
$G_2$ theory not found before in the literature---as well as many new 
isolated solutions (having no marginal deformations).
All our scale-invariant RSK geometries are consistent with an 
interpretation as effective theories of $N=2$ superconformal field 
theories, and, where we can check, turn out to exist as 
quantum field theories.}
 
\begin{document}

\section{Introduction\label{s1}}

Since the pioneering work of Seiberg and Witten \cite{sw94} the rigid 
special Kahler (RSK) geometries of the Coulomb branches of many 
four-dimensional $N=2$ supersymmetric field theories have been found.  
However there is little in the way of systematic understanding of these 
geometries.  One strategy for organizing them is to focus on 
renormalization group fixed points---scale-invariant field theories---and 
then to perturb them by relevant operators.  In this way we 
should get all models on RG flows emanating from a UV fixed 
point.\footnote{This may not include all the possible physical RSK 
geometries; see the discussion in section \ref{s5}.}  In the case of 
a one-complex-dimensional Coulomb branch, the classification of 
scale-invariant RSK geometries is complete \cite{apsw9511,mn96}, and 
is given as a subset of the Kodaira classification of stable 
degenerations of a genus one Riemann surface \cite{kodaira}.   
(We review this classification in section \ref{s2} below.)  
Strikingly, every scale-invariant 1-dimensional RSK 
geometry corresponds to an $N=2$ superconformal field theory (SCFT) 
\cite{apsw9511,excft}.  This suggests the conjecture that all 
scale-invariant RSK geometries correspond to SCFTs.

There is no known classification of higher-dimension scale-invariant
RSK geometries.  Even at (complex) dimension 2, the Coulomb branch geometries 
corresponding to all the classical scale invariant theories (those with 
lagrangian weak coupling limits) are not known.  In particular, the RSK 
geometry for the scale invariant theory with exceptional gauge group 
$G_2$ and four massless hypermultiplets in the 7-dimensional representation 
is not known.  Note that the genus 2 analog of the Kodaira 
classification---all inequivalent, stable, one parameter degenerations 
of a genus 2 Riemann surface---is known \cite{nu73}, but is of little help 
since the RSK geometries of two-dimensional Coulomb branches are described 
by {\em two}-parameter degenerations of genus 2 curves; moreover, not all 
such two-parameter degenerations correspond to RSK geometries.
For higher dimensions, only a few special series of RSK geometries 
corresponding to certain SCFTs realized as limits of lagrangian field
theories, have been catalogued \cite{higherdim}.

In this paper we will introduce a systematic approach to the classification 
of scale-invariant RSK geometries.  Section \ref{s2} starts  with a 
general review of RSK geometry, and its description in terms of a basis 
of holomorphic one-forms on a family of abelian varieties.  
We then re-derive 
the classification of 1-dimensional scale-invariant RSK geometries.
Sections \ref{s3} and \ref{s4} are the heart of the paper, where we 
analyze the 2-dimensional case.  In them we discuss invariances of both 
curve and Coulomb branch under reparametrizations; the formulation of the 
central charge integrability equation; physical monodromy restrictions; 
the scaling assumption; how all these ingredients reduce the classification 
problem to a finite polynomial system; and the consistency between the 
complex singularities and the behavior of BPS masses.  We restrict 
ourselves to discussing only $y^2=x^6$ type singularities (in our gauge 
choice).  Discussion of the other two singularity types ($y^2=x^5$ and 
$y^2=x^3(x-1)^3$) will be left to another paper.  A list of the 
Seiberg-Witten curves for the resulting 12 
scale-invariant RSK geometries appears in table \ref{tab.cft} in section 
\ref{s5}.  In that section we use our solutions to test the hypothesis 
that all scale-invariant RSK geometries correspond to SCFTs, and conclude 
with a discussion of open problems and directions for further work.  
Finally, appendices A and B treat the algebraic details of our solution 
method.


\section{One-dimensional scale-invariant RSK geometries\label{s2}}

We begin with a review of RSK geometry, its connection to $N=2$
supersymmetric effective actions, and its formulation in terms of 
the complex geometry of torus bundles over the Coulomb branch.  
We will then specialize to 1-dimensional RSKs, and reproduce 
the classification \cite{apsw9511,mn96} of all scale-invariant such 
geometries.  Though this material has been understood for many years 
\cite{crtp9703,sw94,dw9510,apsw9511,mn96}, we repeat it here not 
only to set notation, but also to point out some subtleties that 
are important for analyzing higher-dimensional cases.  In particular 
we include a 
discussion of monodromy constraints on the allowed form of the 
Seiberg-Witten curve.

\subsection{Review of RSK geometry}

We define RSK geometry as the geometry of the $r$-complex-dimensional 
Coulomb branch of vacua of the low energy $\U(1)^r$ effective action 
of any $d=4$, $N=2$ supersymmetric field theory.  In particular, if 
$a^i$ is the complex vev of the scalar field in the $i$th $\U(1)$ 
vector multiplet, then supersymmetry implies that the symmetric 
$r\times r$ complex matrix of effective gauge couplings, $\t_{ij}(a)$, 
is a section of an $\Sp(2r,\Z)$ bundle transforming by the standard 
electric-magnetic duality action
\be\label{tausl2z}
\t \to {A\t+B\over C\t+D},
\qquad {A\ B\choose C\ D}\in\Sp(2r,\Z) ;
\ee
is holomorphic in the $a^i$'s; satisfies the integrability 
condition
\be\label{tauint}
{\del\t_{ij}\over\del a^k} - {\del\t_{ik}\over\del a^j} =0 ;
\ee
and determines the hermitean metric on the Coulomb branch $g_{i\bar\jmath}
= \mbox{Im}\,\t_{ij}$, implying that Im$\t_{ij}$ is positive definite.  
Note that these conditions only hold for a particular class of 
coordinates on the Coulomb branch---``special coordinates" 
$a^i$---which are linearly related by supersymmetry to the gauge 
fields.  The integrability condition (\ref{tauint}) implies that 
locally there exist ``dual" complex coordinates $b_i(a)$ on the 
Coulomb branch satisfying
\be\label{tdef1}
\t_{ij} = {\del b_i\over \del a^j} .
\ee
Compatibility with electric-magnetic duality transformations
implies that $(b_i,a^i)$ transforms as a $2r$-component vector
under $\Sp(2r,\Z)$.  In the sector of the theory with magnetic 
and electric charges $(m^i,e_i)$ with respect to the chosen basis 
of $\U(1)^r$, the central charge of the $N=2$ algebra is
\be\label{defZ}
Z = e_i a^i + m^j b_j,
\ee
whose modulus gives a lower bound on the mass of any state with 
those charges.

In what follows, we will work in a simply connected open set $\MM$ 
of the Coulomb branch parametrized by some general (\ie, not 
necessarily ``special") complex coordinates $u^i$.  By definition, 
the $u^i$ are good (single-valued) complex coordinates on $\MM$.
Nevertheless, the effective couplings $\t_{ij}(u)$ may be singular 
on complex subvarieties of $\MM$.  This means, in particular, that 
there can be non-trivial $\Sp(2r,\Z)$ monodromies along paths 
encircling those singularities of complex codimension 1.  We
assume that any singularities in the low energy effective action 
are due to charged $N=2$ multiplets becoming massless.  This 
implies that the BPS masses must vanish at the singularities 
($Z\to0$) since they are the minimum masses for all states in a 
given charge sector.  (Extra neutral states could become massless, 
but they decouple at low energies, and so won't produce a singularity.)

We do not apply any regularity conditions on the metric on $\MM$ 
induced from the scalar kinetic terms.  In section \ref{s4} we 
analyze the nature of the metric singularities for our solutions, 
and find that all metric singularities are integrable (\ie, they 
are at finite distance) in $\MM$.

Following \cite{dw9510}, we can conveniently characterize RSK geometry 
by introducing an auxiliary family of $r$-dimensional abelian varieties 
$T(u^i)$ varying holomorphically over $\MM$.  An $r$-dimensional abelian 
variety is a complex torus $T$ for which there exist $2r$ real 
coordinates $\{x_i,y^i\}$ with $x_i\simeq x_i+1$ and $y^i\simeq 
y^i+1$ the torus identifications, so that the complex coordinates are 
given by $z_i=n_i x_i+\t_{ij}y^j$ for some positive integers $n_i$, and 
with $\t_{ij}=\t_{ji}$ and $\Im\t_{ij}>0$ \cite{gh78}.\footnote{The 
integers $n_i$ are related to the set of electric and magnetic charges 
that appear in the low energy theory \cite{dw9510}.  These integers will 
play no role in what we do: any variety with $N=\prod_i n_i>1$ can be 
obtained as an $N$-sheeted cover of a principally polarized
variety (one with $N=1$); see p.\ 321 of \cite{gh78}.  We will only
deal with principally polarized varieties from now on.}
Then the 2-form defined by $t := dx_i \wedge dy^i$ 
is an integral, closed, positive, (1,1) form.  In coordinate-invariant 
language, a principally polarized abelian variety is a complex torus 
$T$ with a Hodge form $t$ (a closed, integral, positive, (1,1) form) 
satisfying $\int_T t^r = 1$.  Then there exists a basis $\{\a^i,\b_i\}$ 
of homology one-cycles such that $\int_{\a^i\wedge\b_j} t = \d^i_j$. 
Note that such a basis is only specified by this condition up to 
redefinitions ${\a\choose\b}\to M{\a\choose\b}$ for $M\in\Sp(2r,\Z)$.
Then
\be\label{varper}
\t_{ij} = B_{ik} (A^{-1})^k_j, \qquad\qquad 
A^j_k:=\oint_{\a^j}\o_k ,\qquad B_{ik}:=\oint_{\b_i}\o_k,
\ee
where $\{\o_i\}$ is any basis of closed holomorphic (1,0) forms on $T$.  
$\t_{ij}$ is independent of the choice of one-form basis, and 
automatically transforms as (\ref{tausl2z}) under changes of choice 
of canonical basis of one-cycles.  

So far we have not used the integrability condition (\ref{tauint}).
It implies \cite{dw9510} the existence, locally in $\MM$, of a natural 
closed (2,0) form on the total space of the family of abelian varieties 
$T$ fibered over $\MM$, defined by
\be\label{DW2form}
\Omega := dz_i \wedge da^i .
\ee
Here we are taking the special coordinates $a^i$ as our good complex
coordinates around a generic point in $\MM$ where $\t_{ij}$ is
not singular.  $\Omega$ is closed on the total space because  
$d\Omega = (\del dz_i/\del a^j) \wedge da^i \wedge da^j,$ and
$(\del dz_i/ \del a^j) - (\del dz_j/ \del a^i) = 
dy^k [ (\del \t_{ik}/ \del a^j)- (\del \t_{jk}/ \del a^i)] = 0$,
where the last step follows from the integrability condition 
(\ref{tauint}).  In taking the $a^j$ derivatives we held the real 
coordinates $x_i$ and $y^i$ fixed since those are the coordinates
whose periodicity conditions (defining the torus $T$) are 
$a^i$-independent.\footnote{Since $\Omega$ is closed, locally it 
can be written as $d\til\l$ for some meromorphic 1-form $\til\l$
on the total space, whose restriction to $T$ is the Seiberg-Witten 
one-form $\l$.}  An important property of $\Omega$ is that
it has at most one leg in the $T$ fiber directions:  it vanishes
when restricted to $T$ at a fixed point $a^i$ in $\MM$.  (Though
$\Omega$ also has only one leg along $\MM$ in (\ref{DW2form}), this 
is a coordinate-dependent property.)  In \cite{dw9510} it
is shown that the existence of a closed holomorphic two-form with 
this property gives a symplectic structure which associates a 
completely integrable Hamiltonian system to RSK geometries.

The advantage of using the above description of RSK geometries as
abelian varieties fibered over $\MM$ together with the 2-form 
$\Omega$, as opposed to working directly with the $\t_{ij}(a)$, 
is that abelian varieties can be described as the solution of 
complex polynomial equations in projective space.  This allows 
an algebraic treatment which will be essential in what follows.
Furthermore, for dimensions 2 and 3 all principally polarized abelian
varieties can be identified with the Jacobian tori of genus 2 or 3
Riemann surfaces \cite{fk80};  this fact will be exploited in our 
analysis of 2-dimensional RSK geometries in section \ref{s3}.

It is important for our purposes to recast the above description
of RSK geometry in terms of a basis $\{u^i\}$ of good complex
coordinates on all of $\MM$ (and thus, in particular, not the special
coordinates $a^i$), and in terms of the above-mentioned projective
variables instead of the $z_i$'s for the abelian variety.  Noting 
that (\ref{tdef1}) can be written $\t_{ij} = (\del b_i/\del u^k)
(\del u^k/\del a^j)$, we identify $A^j_k$ and $B_{ik}$ in (\ref{varper})
with the derivatives of the special and dual coordinates $a^j$, $b_i$; 
or, more compactly using (\ref{defZ}),
\be\label{BPS}
\del_k Z = \oint_\g \o_k ,
\ee
where $\del_k := \del/\del u^k$, and the homology class of $\g$ 
depends on the charges of the state.  The integrability of (\ref{BPS}) 
then places a restriction on how the basis $\{\o_k\}$ can depend 
on the moduli $u^i$, namely $\oint_\g \del_{[j}\o_{k]}=0$ for all $\g$, 
which in turn implies that
\be\label{integrability}
\del_{[j}\o_{k]}=d_T g_{[jk]}
\ee
for some meromorphic functions $g_{[jk]}$ on the total space of the 
bundle of tori $T(u^i)$ over $\MM$.  Here $d_T$ is the exterior 
derivative along the $T(u^i)$ fibers---\ie, at fixed $u^i$ in the 
total space.  Note that the one-forms $\o_k$ holomorphic on $T$ 
generally are \emph{not} holomorphic as forms on the total space of 
$T$ fibered over $\MM$, but can have poles.  This is why the functions 
$g_{[jk]}$, meromorphic on $T$, can appear on the right hand side of (\ref{integrability}).

It may be helpful to illustrate this last point with a simple example.  
Suppose $T$ is a one-complex-dimensional torus given by the Riemann 
surface $wy^2=x^3 - u^3w^3$ where $(w,x,y)$ and are the homogeneous 
coordinates on $\CP^2$.  $T$ is holomorphically fibered over any set 
$\MM$ in the $u$-plane which does not include $u=0$ or infinity, where 
the fiber degenerates.  The holomorphic one-form on $T$, unique up to 
an overall factor, is $\o = dx/y$, in a coordinate patch with $w=1$.  
At a branch point of the curve for $T$---\eg, $x=u$---the good 
complex coordinate on $T$ is not $x$ but $\xi=\sqrt{x-u}$.  Near 
$\xi=0$, $y\simeq u \xi$ to leading order in $\xi$, and $dx \simeq 
\xi d\xi+du$, so $\o \simeq (d\xi/u) + du/(u\xi)$ which has a pole 
along the $du$ leg at $\xi=0$ for all $u$.

The $\o_i$ can be made into holomorphic (as opposed to meromorphic)
forms $\til\o_i$ on the total space by adding an appropriate one-form 
with legs along the base, $\til\o_i=\o_i+g_{ij}du^j$, where $g_{ij}$
are meromorphic functions.  Then the holomorphic (2,0) form $\Omega 
=\til\o_i \wedge du^i$ is closed if both (\ref{integrability}) is 
satisfied and also  
\be\label{DW2form2}
\del_{[i}g_{jk]}=0.
\ee
In what follows, our strategy for solving for possible RSK 
geometries will be to impose (\ref{integrability}) and (\ref{DW2form2})
which are the translation into general projective coordinates of the 
original RSK integrability condition (\ref{tauint}).

There is a further condition on the $g_{[ij]}$, namely that the closed 
(2,0) form $\Omega$ be holomorphic (as opposed to meromorphic) in the total
space over any open set in $\MM$ where the fibers do not degenerate. 
Imposing the holomorphy of $\Omega$ from the beginning is difficult; 
however, after picking an explicit basis of $\o_i$ (in section 3.6), 
we will be able to show that the integrability condition 
(\ref{integrability}) automatically implies the holomorphy of $\Omega$ 
for two-dimensional RSK geometries.

To summarize, an $r$-dimensional RSK geometry can be characterized
by a holomorphic $T$ bundle over an open set $\MM \subset \C^r$,
together with a basis of closed holomorphic one-forms $\{\o_i\}$ 
on the fibers satisfying the central charge integrability condition 
(\ref{integrability}) and the RSK integrability condition 
(\ref{DW2form2}).  Note that when the Coulomb branch is just one- 
or two-dimensional---as is the case in this paper---(\ref{DW2form2})
is trivially satisfied, and only (\ref{integrability}) needs to
be examined.

\subsection{Classifying dimension-1 scale-invariant RSK geometries}

We now specialize to one-dimensional RSK geometries.  A complex torus 
$T^2$ is conveniently described as an algebraic curve.
There are many ways of doing this.  The most familiar is as the zeros of 
a degree 3 polynomial in $\CP^2$; another is the hyperelliptic form, 
as the zeros of a degree 4 polynomial in the weighted projective space 
$\CP^2_{(1,1,2)}$.  We will classify the scale-invariant RSK geometries
using both descriptions in parallel because this will allow us to illustrate
some points which will be important in the two-dimensional case.

For both descriptions of the torus, let $(w,x,y)$ be homogeneous coordinates
on the projective space.  Some of the freedom of holomorphic coordinate
redefinitions of $(w,x,y)$ can be used to put a general elliptic curve
and holomorphic one form in the convenient canonical forms (written in the 
$w=1$ coordinate patch):
\bea\label{crvs2}
y^2 &=& x^4 + a x + b,
\qquad \o=dx/y, \qquad\mbox{(hyperelliptic)},\nonumber\\
y^2 &=& x^3 + c x + d,
\qquad \o=dx/y, \qquad\mbox{(cubic)} .
\eea
We are also free to make holomorphic reparametrizations of the 
complex coordinate $u$ on the Coulomb branch: $u\to \til u(u)$.
We will return to this freedom a bit later, after we have imposed
the scaling conditions.

We make a holomorphic $T^2$ bundle over $\MM$ by allowing all the
coefficients in the above equations to vary holomorphically in $u$.  
So, to classify one-dimensional RSK geometries, we want to solve 
for $a,b,c,d$ as functions of $u$ compatible with (\ref{integrability}).  
Since we have only a single coordinate $u$ on $\MM$, the integrability 
condition (\ref{integrability}) is empty.  

The interesting restrictions come from demanding consistent global 
continuations in $u$, obeying some physical boundary conditions as 
$u\to\infty$ and as $u$ approaches values at which the complex 
structure of the torus degenerates.  As explained in the introduction, 
we will focus here on the boundary conditions near singularities.  
The basic physical insight is that a singularity in the low energy 
Wilsonian effective action of a well-defined field theory must be a 
reflection of new charged degrees of freedom becoming massless.  Choose 
coordinates $u^i$ so that the singularity in question is at the origin.  
So, as we approach the singularity, the low energy effective action will 
be dominated by the behavior of these new light degrees of freedom whose 
mass vanishes as some positive power of $|u^i|$.  (Holomorphy disallows 
the central charge from varying exponentially with $u$, since this gives 
an essential singularity at $u^i=0$, and so there would be some directions 
in which masses would diverge.)  Thus, by going close enough to the 
singularity, the RSK geometry should be dominated by some uniform scaling.  
In particular, approaching $u^i=0$ along a path with all $u^i=0$ for 
$i\neq j$, the BPS mass $|Z|$ of the lightest state vanishes as some 
power of $u^j$, or, inverting the relationship,
\be\label{rank1bb}
u^j \sim Z^{[u^j]},
\ee
which serves to define the scaling, or mass, dimension of $u^j$.
We will always denote scaling dimensions by square brackets.

Note that the scaling dimensions need not be real.  For instance,
discretely self-similar scaling behavior, corresponding to a limit 
cycle in the RG flow, would manifest itself as complex scaling 
dimensions.  Nevertheless, they must always have positive real parts 
in order for the masses to vanish at $u^i=0$: this simply reflects 
our assumption that that is where the singularity is.
In summary, to be as inclusive as possible in our classification, 
we only demand of our scaling exponents that their real parts be
positive:
\be\label{posexp}
\Re ([u^j])>0.
\ee

In the 1-dimensional case this means that as $u\to0$, the
curves (\ref{crvs2}) should become singular, corresponding to 
branch points (zeros of the right-hand sides) colliding.
The maximal number of colliding branch points in this presentation
of the curve is 4 or 3 at $x=0$ in the hyperelliptic case or 
the cubic case, respectively.  Thus, at $u=0$ the
curves will degenerate to $y^2=x^4$ or $y^2=x^3$, respectively.
This happens when $a,b,c,d$ all scale as $u$ to some powers with 
positive real parts, 
\bea\label{crvs3}
y^2 &=& x^4 + a u^\a x + b u^\b,\qquad \Re\a,\ \Re\b>0,
\qquad \qquad\mbox{(hyperelliptic)},\nonumber\\
y^2 &=& x^3 + c u^\g x + d u^\d,\qquad\, \Re\g,\ \Re\d>0,
\qquad \qquad\ \mbox{(cubic)},
\eea
where now $a,b,c,d$ are complex constants.  Although we have
written only a single power of $u$ in each term, we do not
make that assumption in what follows.  The power with the smallest 
real part will dominate the scaling behavior as $u\to0$.  But,
since we allow complex exponents, there may be terms with
different imaginary parts in the exponents but all with the
same real part.  Alternatively, an accumulation point of real exponents
may be allowed, in which case just keeping the lowest exponent
may not be accurate.

In addition to the $y^2=x^4$ or $y^2=x^3$ degenerations assumed
in (\ref{crvs3}), there can also be sub-leading
degenerations, where fewer than the maximal number of
branch points collide.  We will examine those singularities 
later in this section.

We have not used our freedom to make holomorphic reparametrizations
on our patch $\MM$ of the Coulomb branch.  The scaling assumption,
however, tightly constrains this freedom.  First, we want to
keep the singularity at $u=0$, so only holomorphic reparametrizations
of the form $u\to f u + g u^2 + \cdots$ with $f$ a non-zero complex
constant are allowed.  But then, upon scaling to $u=0$, the higher-power 
terms will all be subleading, and so can be dropped.  Thus
the only reparametrization compatible with scaling behavior is
an overall rescaling of $u$.  The BPS equation (\ref{BPS}), which 
reads here
\be\label{rank1b}
\del_u Z = \oint dx/y ,
\ee
implies that a rescaling of $u$ is accompanied by a rescaling of
the one-form $dx/y$.  The normalization of the one-form and of
the leading terms in the curves (\ref{crvs3}) can be maintained
by appropriate rescalings of $x$ and $y$: $(u,x,y)\to(f u,f x,
f^2 y)$ for the hyperelliptic curve, and $(u,x,y)\to(f u,f^2 x,
f^3 y)$ for the cubic curve.  These, in turn, imply that the 
coefficients of the curves are rescaled by 
\be\label{rank1a}
a\to f^{\a-3} a, \qquad
b\to f^{\b-4} b, \qquad
c\to f^{\g-4} c, \qquad
d\to f^{\d-6} d.
\ee
This freedom allows us to rescale one coefficient in each
curve to $1$ if it does not vanish. We will only know which
coefficients vanish upon completing the classification of
scale-invariant curves, so we will use this freedom at the
end.

The scaling assumption also implies constraints among the exponents 
of separate terms.  For uniform scaling behavior, $x$ and $y$ must 
also scale with some definite dimensions.  Then (\ref{rank1b}) implies
the relation among scaling dimensions
\be\label{rank1cc}
1-[u]=[x]-[y].
\ee
Also, each term of the curves (\ref{crvs3}) must scale in the 
same way, implying
\bea\label{rank1c}
2[y] &=& 4[x]=\a [u]+[x]=\b [u],
\qquad\mbox{(hyperelliptic)},\nonumber\\
2[y] &=& 3[x]=\g [u]+[x]=\d [u],
\qquad\, \mbox{(cubic)}.
\eea
(\ref{rank1cc}) and (\ref{rank1c}) can be solved to give
\bea\label{rank1e}
{} [u] &=& \frac{3}{3-\a} = \frac{4}{4-\b},
\qquad\mbox{(hyperelliptic)},\nonumber\\
{} [u] &=& \frac{4}{4-\g} = \frac{6}{6-\d},
\qquad\, \mbox{(cubic)}.
\eea

So far our scaling assumption only requires that the real
part of $[u], \a, \b, \g$, and $\d$ be positive, which
leaves a continuous range of possible scaling
solutions.  However, we must also demand that the
low energy physics be single-valued in $\MM$.
By assumption, $u$ is a good coordinate on $\MM$, so
upon letting $u$ traverse any path encircling the origin,
\eg\ $u \to e^{2\pi i}u$, the low energy data---the
complex structure of the curve, $\t$, and the central
charge, $Z$---must return to their same values modulo 
$\Sp(2,\Z)$ transformations.  This implies
that the holomorphic one-form and the form of the curve
must remain invariant, up to possible changes of variables
of $x$ and $y$.  The only such globally holomorphic changes
of variables compatible with scaling are
$x \to x e^{i \x}$, and $y \to y e^{i \eta}$.
Invariance of the BPS equation (\ref{rank1b}) implies that 
$e^{i\x-i\eta}=1$.  Comparing to the form of the curves 
(\ref{crvs3}) then gives $1 = e^{2i\eta} = e^{2\pi i\a-i\eta}
= e^{2\pi i\b-2i\eta}$ in the hyperelliptic description, and
$1 = e^{i\eta} = e^{2\pi i\g-i\eta}=e^{2\pi i\d-2i\eta}$ in the
cubic description.  The only solutions are 
\bea\label{rank1f}
\a\in\Z/2\ &\mbox{and}&\ \b\in\Z,
\qquad\ \mbox{(hyperelliptic)},\nonumber\\
\g\in\Z\ &\mbox{and}&\ \d\in\Z,\ 
\qquad\mbox{(cubic)}.
\eea
Thus the allowed powers of $u$ appearing in the curve are real and
discrete.  

\TABLE[ht]{
\begin{tabular}{||c|r|c||}   \hline
Dimension & \multicolumn{1}{c|}{Cubic curve} & Global symmetry\\
$[u]=\cdots$ & $y^2=x^3+\cdots$ & $\U(2)_R\times\cdots$\\ \hline
6/5			& $u\,\,$		& $\U(1)$	\\ 
4/3			& $u\,x\qquad\ \ \,$& $\U(2)$	\\ 
3/2			& $u^2$			& $\U(3)$	\\ 
2			& $u^2x+\t u^3$	& $\SO(8)$	\\ 
3			& $u^4$			& $E_6$		\\ 
4			& $u^3x\qquad\ \ \,$& $E_7$		\\ 
6			& $u^5$			& $E_8$		\\ \hline
\end{tabular}
\caption{Dimension-1 scale-invariant solutions (in cubic form) 
with their global symmetry groups.}\label{tab2}}

Combining the constraints (\ref{rank1e}, \ref{rank1f}), and
Re$(\a,\b,\g,\d)>0$ gives a finite list of scaling solutions.
Note that the set of solutions in the hyperelliptic and cubic
descriptions of the curve coincide, as they must.  
Table \ref{tab2} gives the curves corresponding to
these solutions in their cubic description.  We have used the
$u$ reparametrization freedom (\ref{rank1a}) to set
one (subleading) coefficient to 1 in each curve.  The
remaining unfixed coefficient $\t$ in the $\SO(8)$ curve
is thus an exactly marginal deformation of the associated
$N=2$ theory: indeed, it is the marginal gauge coupling of
the scale-invariant $SU(2)$ theory with either four massless
fundamentals, or one massless adjoint \cite{sw94,apsw9511}.  
(We have also included for reference purposes the global 
flavor symmetry groups associated to the mass deformations 
of these theories \cite{apsw9511,mn96}.)

Note that in all these cases $[u]>1$, though this was not
required \emph{a priori}.  This is consistent with the assumption 
of conformal 
invariance which implies that $[u]>1$ for any operator in an 
interacting conformal field theory (CFT), $[u]=1$ for a free 
theory, and $1>[u]>0$ is disallowed by unitarity \cite{scft}.  
This is in line with the expectation \cite{pol88} that the IR 
scaling behavior of local field theories should correspond to 
a CFT.

In fact, all the curves in table \ref{tab2} have been found
to describe superconformal field theories.  Referring to them
by their global groups, the $\U(1)$, $\U(2)$ and $\U(3)$
curves were found , \eg, by tuning one, two or three
fundamental masses in the $\SU(2)$ $N=2$ gauge theory
\cite{apsw9511}, while conformal field theories corresponding 
to the $E_n$ curves were found by compactifying certain 6- or 
5-dimensional field theories constructed using string techniques
\cite{excft}.  
It is striking that all 1-dimensional scale-invariant RSK
geometries are realized as the Coulomb branches of $N=2$ quantum
field theories, and suggests the hypothesis that all scale-invariant
RSK geometries are so realized.  Testing this hypothesis is a
large part of the motivation for the analysis of 2-dimensional
scale-invariant RSK geometries undertaken in this paper.

\subsection{Sub-leading singularities in 1-dimensional RSK geometries}

So far we have only analyzed the maximal singularity where
all three or four branch points coincide for the cubic or
hyperelliptic form of the curves, respectively.  There are
also sub-leading singularities, where fewer numbers of
branch points collide.  Let's analyze these for the cubic 
curve; we'll check that the hyperelliptic form gives the same
results later.  For the cubic curve, the only 
possibility is a ``(2,1) singularity", where two branch points
collide while the third remains at a distinct point:
\be\label{subcrv1}
y^2 = x^3 - 2e^2 x + 3e^3 = (x-e)^2 (x+2e).
\ee
(We could use the rescaling freedom (\ref{rank1a}) to set $e$ to any
convenient nonzero value, say $e=1$, if we wanted to; however, it
will be convenient to keep it unfixed because it will be interpreted
as a dimensionful parameter.)
If, by assumption, we approach the above singularity as $u\to0$,
then states with charges corresponding to a cycle enclosing just the
two branch points near $x=e$ will give vanishing masses.  From
(\ref{rank1b}), $Z$ will be dominated by the contribution from
$x\sim e$.  Thus we can take the scaling limit by changing variables 
to $\x=x-e$ in (\ref{subcrv1}), adding generic $u$-dependent deformations
as in (\ref{crvs3}), and keeping only the leading (lowest)
powers of $\x$:
\be\label{subcrv2}
y^2 = \OO(\x^3) + 3e \x^2 + c u^{\g} \x + d u^{\d}.
\ee
Then scaling of (\ref{rank1b}) implies the relations among scaling
dimensions $1-[u]=[\x]-[y]$ and $2[y] = 2[\x]= \g [u]+[\x] = \d [u]$,  
whose only solutions are $[u]=1$ and $\d=2\g$.  The monodromy
argument leading to (\ref{rank1f}) still applies, so that
there is an infinite series of possible singularities,
\be\label{subcrv3}
y^2 = (x-e)^2 (x+2e) + c(x-e) u^n + d u^{2n},
\qquad\qquad n\in\Z>0.
\ee
(There is also a series of solutions with $c=0$ and
$\ldots+d u^n$.)  Since these
all have $[u]=1$, we expect them to correspond to free CFTs.
Indeed, from the monodromies around the singularities \cite{sw94},
these are easily recognized as the IR free $\U(1)$ theories 
corresponding to $n$ particles with the same charge becoming 
massless.  
 
Since $[\x]=[u^n]$ and $[u]=1$, $\x\sim \mbox{(mass)}^n$; so we 
should assign $x$ and $e$
dimensions of (mass)$^n$.  Because of this fixed mass scale, the curve 
(\ref{subcrv3}) cannot describe a true scale-invariant theory.  
Specifically, the masses of states with charges corresponding to 
cycles which are not shrinking (\eg, one encircling $x=e$ and $x=-2e$)
will depend on the mass parameter $e$; whereas the masses of
the states corresponding to the vanishing cycle are independent of
$e$ in the $u\to0$ limit, as can easily be checked explicitly from
(\ref{rank1b}).  Furthermore, the vanishing masses are also
independent of the parameters $c$ and $d$ in the curve.  This can
be seen from the form of the curve (\ref{subcrv2}) near the singular 
locus:  rescalings of $y$, $\x$, and $u$, and shifts in $\x$ can
bring the singularity locally to the form $y^2 \sim \x^2 + u^n$.

An analogous story holds for the hyperelliptic form of the curve
(as it must).  The only novelty is that for the hyperelliptic
form of the curve, where there are four branch points, there
are three distinct subleading singularities: those of type (3,1),
where 3 roots collide while the fourth stays separate; type (2,2), 
where 2 pairs separately collide; and (2,1,1), where only a single 
pair collides.  However, we must also demand that these singularities
be consistent with the chosen form (\ref{crvs2}) of the curve.
It is not hard to see that the (2,1,1) singularity is the only
consistent one, giving the singular curve $y^2 = x^4 - 4e^3 x + 3e^4 
= (x-e)^2 (x^2+2ex+3e^2)$.
Then, scaling $\x=x-e\to0$ with general $u$-dependent deformations,
$y^2 = \OO(\x^3) + 6e^2 \x^2 + a u^{\a} \x + b u^{\b}$, gives the
relations $1-[u]=[\x]-[y]$ and $2[y] = 2[\x]= \a [u]+[\x] = \b [u]$.
The only solutions to these equations are $[u]=1$ and $\b=2\a$ leading
to a series of singularities $y^2 = (x-e)^2 (x^2+2ex+3e^2) + 
a(x-e) u^{n/2} + b u^n$, $n\in\Z>0$, corresponding to IR free $\U(1)$ 
theories with $n$ massless particles of the same charge.


\section{Two-dimensional scale-invariant RSK geometries\label{s3}}

As reviewed in the last section, we can describe 2-dimensional 
RSK geometries on a simply connected open set $\MM\subset \C^2$ 
in terms of a family of 2-dimensional principally polarized abelian 
varieties together with a choice of basis of holomorphic one-forms 
varying holomorphically over $\MM$.  Such varieties can all be 
identified with Jacobian tori of genus 2 Riemann surfaces \cite{fk80}.  
This permits the economical algebraic description of these spaces 
as a family of algebraic curves varying over $\MM$.   The conditions 
on the holomorphic one-forms on the tori described in section 
\ref{s2}.1 descend trivially to the Riemann surface: equations 
(\ref{varper}) and (\ref{integrability})--(\ref{DW2form2}) remain 
unchanged for $\{\o_k\}$ a basis on holomorphic one-forms on the 
Riemann surface.  (The $\a^i$ and $\b_j$ cycles become a basis of 
homology cycles on the Riemann surface with intersections $\a^i\cdot
\b_j=\d^i_j$ and $\a^i\cdot\a^j=\b_i\cdot\b_j=0$.)

\subsection{Curve and forms}

All genus 2 curves are hyperelliptic, and so can be written as a 
degree 6 polynomial in $\CP^2_{(1,1,3)}$ \cite{fk80}.  In 
homogeneous variables $(w,x,y)$, using holomorphic reparametrizations 
to shift away terms linear in $y$, and in the $w=1$ coordinate patch, 
the general curve can be written as
\be\label{crv}
y^2 = \sum_{k=0}^6 x^k c_k(u,v).
\ee
Here $u,v$ are good complex coordinates on $\MM$.
Our goal is to determine the allowed forms of the $c_k(u,v)$ coefficients.
We simplify the problem by just looking for the form of the curve in
the vicinity of a degeneration---a locus in $\{u,v\}$ where two or more
zeros in $x$ of the right side of (\ref{crv}) coincide.  We work in an open 
set $\MM\subset\C^2$ around $u=v=0$, where we assume the curve 
degenerates.  By definition, the coordinates $u$ and $v$ are taken to 
be single-valued complex coordinates on this patch of the moduli space.  

These coordinates are tied to properties of the curve by the BPS 
equations (\ref{BPS}) which state that the central charge $Z$ in a sector 
of the theory with given electric and magnetic charges satisfies
\be\label{genbps}
\del_u Z = \oint \til\o_0,\qquad
\del_v Z = \oint \til\o_1,
\ee
where the path of the integration depends on the charges of the state, and
$\{\til\o_0,\til\o_1\}$ form some basis of holomorphic 1-forms on the curve.  
Thus, any basis can locally be written as
\be\label{genbas}
\pmatrix{\til\o_0\cr \til\o_1} =
\pmatrix{\AA&\BB\cr \CC&\DD}
\pmatrix{\o_0\cr\o_1} ,
\ee
where ${\AA\ \BB\choose \CC\ \DD}\in\GL(2,\C)$ with entries holomorphic in 
$(u,v)$, and $(\o_0,\o_1)$ is a basis of holomorphic 
1-forms which we conventionally take to be
\be\label{forms}
\o_0 := {xdx\over y}, \qquad \o_1 := {dx\over y}.
\ee

The curve (\ref{crv}) and basis of holomorphic 1-forms (\ref{genbas}) 
completely specify the RSK geometry on $\MM$.  This description 
is highly redundant, however, and it behooves us to eliminate this 
redundancy through some conventional choices of the structures entering 
in (\ref{crv})--(\ref{genbas}).

\subsection{Reparametrization invariance}

We are free to redefine the variables $x$, $y$ we use to describe the 
curve.  In particular, the transformations of $x$ and $y$ which preserve 
the algebraic form of (\ref{crv}) are
\be\label{redef}
x \to {\til \AA x+\til\BB\over\til\CC x+\til\DD},\qquad
y \to {(\til\AA\til\DD-\til\BB\til\CC)\over(\til\CC x+\til\DD)^3}\, y,
\ee
for ${\til\AA\ \til\BB\choose \til\CC\ \til\DD}\in\GL(2,\C)$ and locally
holomorphic in $(u,v)$.  These transform the conventional 1-forms by
\be\label{redef2}
\pmatrix{\o_0\cr \o_1} \to
\pmatrix{\til\AA&\til\BB\cr \til\CC&\til\DD}
\pmatrix{\o_0\cr\o_1}.
\ee
By choosing ${\til\AA\ \til\BB\choose \til\CC\ \til\DD} 
= {\AA\ \BB\choose \CC\ \DD}^{-1}$ we completely fix this redundancy, 
setting the basis of 1-forms on the curve to be the canonical basis 
(\ref{forms}), so that the BPS equations read
\be\label{bps}
\del_u Z = \oint {xdx\over y},\qquad
\del_v Z = \oint {dx\over y}.
\ee
Note that this is qualitatively different from the genus 1 case.  There
the redefinitions of $x$ and $y$ not only enabled us to fix the form
of the one-form basis, but also to fix all but two coefficients of the
curve.  Here, though, the transformations (\ref{redef}) 
are just enough to fix the basis of one-forms, and
put no further restrictions on the coefficients of the curve (\ref{crv}).

We still have the freedom to redefine the $u$ and $v$ coordinates on 
$\MM$, $\{u,v\}\to\{\til u,\til v\}$ with
\be\label{cov}
u=u(\til u,\til v), \qquad
v=v(\til u, \til v).
\ee
Since they are to be good complex coordinates on $\MM$, and we have fixed 
$u=v=0$ to be the singular point, (\ref{cov}) must obey $u=v=0$ when
$\til u=\til v=0$, and be invertible in an open set around the origin.
The transformation of the $u$ and $v$ derivatives under this 
change of variables will induce \emph{via} the BPS equation (\ref{bps}) a
linear redefinition of the holomorphic 1-forms.  This, in turn, can be 
cancelled by the appropriate $u$- and $v$-dependent redefinition of $x$ 
and $y$ of the form (\ref{redef}) with
\be\label{GLconst}
\pmatrix{\til\AA & \til\BB\cr \til\CC & \til\DD}=
\pmatrix{\del_{\til u}u&\del_{\til u}v\cr \del_{\til v}u&\del_{\til v}v}.
\ee
So---just to be clear---any transformation taking
\be\label{reparam}
u \to \til u,
\qquad
v \to \til v,
\qquad
x \to {x\del_{\til u}u+\del_{\til u}v\over x\del_{\til v}u+\del_{\til v}v},
\qquad
y\to {\del_{\til u}u\del_{\til v}v-\del_{\til u}v\del_{\til v}u
\over(x\del_{\til v}u+\del_{\til v}v)^3}y,
\ee
will leave the canonical basis of 1-forms (\ref{forms}) unchanged, 
but will take the curve (\ref{crv}) to a new curve of the same form, 
but with different coefficients $c_k$.  We will refer to this redundancy 
in our description of the curve as its ``holomorphic reparametrization 
invariance.''

\subsection{Singularity types}

Since a holomorphic reparametrization acts by a fractional linear 
transformation on $x$, it can be used to send any three distinct 
points on the $x$ projective plane to any (distinct) chosen values.  
The natural points to fix in this way are the branch points of the 
curve (\ref{crv}).  But, even though the $u$- and $v$-dependent
parameters in this transformation depend on two functions, it is 
important to note that they generally \emph{cannot} be used to fix 
a $u$- and $v$-independent position of any branch point in the 
$x$-plane.  This is 
because the positions of the branch points are not single-valued 
functions of $u$ and $v$---typically closed paths in $u$ and $v$ 
interchange the various branch points---while the reparametrization 
functions (\ref{cov}) are constrained to be single-valued.  We can, 
however, certainly fix 3 of the distinct branch point positions at 
any single point in $\MM$.  The natural place to impose this choice 
is at $u=v=0$, where the curve is singular.  The singularity can be 
viewed as branch points of $y$ in (\ref{crv}) colliding in the 
$x$-plane.  So we can use the holomorphic reparametrization invariance 
to fix three of the collision points to some conventional values.  

Unlike the genus-1 case, now there are many possibilities compatible
with the form of the curve (\ref{crv}) for how the branch points
can collide.  They correspond to all the ways of partitioning
the six branch points into subsets that collide: (6), (5,1),
(4,2), (4,1,1), (3,3), (3,2,1), (3,1,1,1), (2,2,2), (2,2,1,1),
and (2,1,1,1,1), in the notation used in section \ref{s2}.3.  
Of these, the (6), (5,1), and (3,3) singularities are
special since every cycle on the Riemann surface is homologous
to a vanishing cycle.  As we saw at the end of section \ref{s2},
this means that these types of singularities can correspond to
scale-invariant RSK geometries.  All the other singularity types
will correspond to sub-leading singularities, where only the states
in a subset of the charge sectors will become massless, while states
in the other sectors remain massive as the singularity is approached.

In this paper we will further specialize our problem by concentrating 
on a particular form of the singularity, namely the type (6), where all 
six branch points collide at a single point in the $x$-plane.  We will
examine the other singularity types elsewhere.  As we will see in
section \ref{s5}, the other singularity types are just as important
as the type (6) ones, and include many scale-invariant RSK geometries;
our classification will be extended to include the (5,1) and (3,3) 
singularity types in a forthcoming paper.

We can use the holomorphic reparametrization invariance to choose the 
point where all six branch points collide to be $x=0$, so that 
as $u,v\to0$, the curve approaches the form $y^2\sim x^6$.  This means, 
in particular, that the coefficient of the $x^6$ term is not identically 
zero, allowing us to write the curve as
\be\label{crv6}
y^2= a(u,v)\,f(u,v,x)=a(u,v)\,\left(x^6+\sum_{k=0}^5 f_k(u,v)\, x^k\right).
\ee
Also, because we are assuming a $y^2\sim x^6$ singularity at $u=v=0$, the 
$f_k$ are required to vanish there.  

\subsection{Absence of monodromy in physical quantities}

We now show that the $f_k(u,v)$ must be single-valued functions 
because $u$ and $v$ are assumed to be good local coordinates on 
the moduli space, \ie, all physical quantities are single-valued as 
functions of $u$ and $v$.  (This is the analog of the monodromy
argument that fixed the exponents of $u$ to be integer or half-integer
in the genus 1 case.)  Since the complex structure of the curve 
(modulo choice of canonical homology basis) computes the effective 
coupling (modulo electric-magnetic duality),  it should be single-valued 
as a function of  $u$ and $v$.  Thus, in particular, as we take $u\to 
e^{2m\pi i}u$ and $v\to e^{2n\pi i}v$ along any closed contour in $\MM$, 
the curve (\ref{crv6}) must come back to a curve which has the same 
form up to a possible redefinition of $x$ and $y$ of the form (\ref{redef}).
Since the absolute values of the periods of the 1-forms (\ref{forms}) 
compute the derivatives of BPS masses, under the above monodromies and after 
the change of $x$- and $y$-variables, the basis of 1-forms should remain 
unaltered up to a possible multiplication by an overall ($u$- and 
$v$-independent) phase.  But a general redefinition of $x$ and $y$, 
(\ref{redef}), changes the basis of 1-forms by (\ref{redef2}), so the
only redefinitions of $x$ and $y$ allowed by invariance of the 1-form 
basis are phase rotations.

Now consider any single term in the curve, \eg, one found by expanding 
$f_k$ around $u=v=0$:
\be\label{crvi}
y^2 = a\left(x^6 +\ldots+ u^b v^c x^k +\ldots\right).
\ee
Under $u\to e^{2m\pi i}u$ and $v\to e^{2n\pi i}v$ this becomes
\be
y^2 = e^{i\a_{mn}}a\left(x^6 + \ldots + e^{2\pi i (mb+nc)}
u^b v^c x^k +\ldots\right),
\ee
where $\a_{mn}$ is whatever phase that $a(u,v)$ picks up.\footnote{If $a$ 
were to change by something other than a phase, then phase rotations of $x$ 
and $y$ would not return us to the original curve.  We show on physical 
grounds in the discussion at the end of section \ref{s4}.1 why $a$ can, 
at most, transform only by a phase under any monomdromy in $u$ and $v$.}  To 
put the first term back in the form (\ref{crvi}), we 
redefine $x$ and $y$ by the phases $y \to e^{i(3\b_{mn}+ \a_{mn}/2)}y$, 
$x\to e^{i\b_{mn}}x$, where $\b_{mn}$ is arbitrary.  To put the other 
terms back to the form (\ref{crvi}) we need $(6-k)\b_{mn}=2\pi(mb+nc)$ 
(mod $2\pi$).  But the ratio of the 1-forms $(\o_u/\o_v)$ picks up the 
same phase $\b_{mn}$ as $x$, so to be invariant, we must have $\b_{mn}=0$ 
(mod $2\pi$) for all $m,n$.  Taking $\{m,n\}=\{1,0\}$ and $\{0,1\}$, 
implies that $b,c\in\Z$, showing that the 
$f_k(u,v)$ are single-valued in $u$ and $v$.  Note that there is no 
such restriction on the prefactor $a(u,v)$.

\subsection{Integrability of the BPS equations}

The most important constraint on the form of the curve (\ref{crv6}) comes 
from the BPS equations (\ref{bps}).  A necessary condition for the 
existence of a solution to the BPS equations is the (local) integrability 
condition
\be
0=\del_u\del_vZ-\del_v\del_uZ =
\del_u\oint {dx\over y} -\del_v \oint{xdx\over y} .
\ee
Since this should hold for arbitrary integration path on the curve, it
follows that the integrand must be a total derivative.  Furthermore, since
the integrand is odd under the $\Z_2$ automorphism of the curve taking
$(x,y)\to(x,-y)$, it follows that we can take the total derivative also
to be odd, giving the consistency condition
\be\label{rsk}
\del_u \left({1\over y}\right) 
-\del_v \left({x\over y}\right)
= \del_x \left({b\over y}\right),
\ee
where $b(u,v,x)$ must be regular (single-valued) in $x$ to be 
well-defined on the curve, but is otherwise arbitrary.  The condition 
(\ref{rsk}) has no analog in the genus-1 case.  It is the translation
to projective coordinates of (\ref{integrability}) from our general 
discussion of RSK geometries in section 2.1.  Equation (\ref{rsk})
will play a central role in the analysis of this section.  We will
refer to it as the ``integrability equation" in the discussions to 
follow.

We now show that $b$ must, in fact, be at most a 3rd-order polynomial in 
$x$ with arbitrary $u,v$-dependence.  We can see this by taking various 
limits in $x$ of (\ref{rsk}).   In the limit $x\to x_0$, $x_0$ any finite 
point, the leading $(x-x_0)$-dependence of $y$ is  $y\sim (x-x_0)^p$ for 
some $p\ge0$; typically, $p=0$.  Thus $(\del_u-x\del_v) y^{-1}\sim 
(x-x_0)^{-p+n}$ for $n\ge 0$ in this limit; typically, $n=0$, unless 
there is some cancellation.  Now suppose that in this limit $b\sim 
(x-x_0)^m$ for some $m$.  Then the right side of (\ref{rsk}) goes as ${}\sim 
(x-x_0)^{m-p-1}$ for $m\neq p$, and ${}\sim (x-x_0)^\ell$ with $\ell\ge0$
for $m=p$, since in the latter case the leading $(x-x_0)$-dependence of 
$b$ and $y$ cancel, so that subleading terms contribute.  Equating left 
and right sides implies $m\ge1$ for $m\neq p$, or $m=p\ge0$.  
Therefore, we have determined that $b(x)$ is regular at all finite $x$.  
As $x\to\infty$, $y\sim x^3$, so $(\del_u - x\del_v) y^{-1}\sim 
(x-x_0)^{-2+n}$ for $n\ge 0$; typically, $n=0$, unless there 
is some cancellation.  Now suppose that in this limit $b\sim x^m$.  
Then the right side of (\ref{rsk}) goes as ${}\sim x^{m-4}$ 
for $m\neq 3$, and ${}\sim x^{-\ell}$ with $\ell\ge0$ for $m=3$, since 
in the latter case the leading $x$-dependence of $b$ and $y$ cancel, so 
that subleading terms contribute.  Equating left and right sides implies 
$m\le2$ for $m\neq 3$, or $m=3$.  Therefore, we have determined that 
$b(x)$ cannot grow faster than $x^3$ at infinity.   Altogether this implies 
that $b$ is a polynomial of order 3 in $x$.

Expand (\ref{rsk}) using (\ref{crv6}) in terms of $a(u,v)$, $b(u,v,x)$, 
and $f(u,v,x)$ as
\be\label{rski}
f[\del_u-x\del_v] \ln a
+[\del_u-x\del_v]f
= b\del_xf-2f\del_xb .
\ee
Parametrize the polynomial $x$-dependence of $f$ and $b$ by
\be\label{effexp}
f = x^6 + \sum_{k=0}^5 x^k f_k,
\qquad
b = \sum_{a=0}^3 x^a b_a,
\ee
and expand (\ref{rski}) out order-by-order in $x$.  Only the powers $x^0$
through $x^8$ contribute.  It is a short exercise to show that the $x^8$
equation is identically satisfied, and that the $x^{7,6}$ equations 
determine $\del_{v,u}\ln a$ as polynomials in $f_k$, $\del_{u,v}f_k$, 
and $b_a$.  Furthermore, $\del_{u,v}\ln a$ is linear in $b_a$. 
Substituting these expressions back into (\ref{rski}) to eliminate
$\del_{u,v}\ln a$ gives a set of 6 equations (the coefficients of $x^0$ 
through $x^5$) which are linear in $b_a$ and polynomial in $f_k$ and 
$\del_{u,v}f_k$.  The equations can be solved for the $b_a$, leaving us 
with two coupled non-linear partial differential equations in $u$ and 
$v$ for the six $f_k(u,v)$.  So far, the $f_k$ are only constrained to 
be regular in $u$ and $v$ and to vanish at $u=v=0$; they also have a 
large redundancy under holomorphic reparametrizations (\ref{reparam}).

\subsection{Holomorphy of $\Omega$}

We now show that the integrability equation (\ref{rsk}) implies
the holomorphy (for generic $u$ and $v$) of the (2,0) form $\Omega$ 
introduced in section 2.1.  There the integrability equation 
(\ref{integrability}) took the form $\del_{[u}\o_{v]} = \del_x g_{[uv]} 
dx$.  Comparing to (\ref{rsk}) implies $g_{[uv]}=b/y$.  The 2 form is 
\be\label{Omega1}
\Omega := \o_u\wedge du + \o_v\wedge dv + g_{[uv]} du\wedge dv
= y^{-1} \left[dx\wedge (x du+dv) + b\, du\wedge dv\right].
\ee
The possible poles in $\Omega$ are at $y=0$, which are the
branch points of the curve, where $x=r(u,v)$ such that $f|_{x=r}=0$.
For generic $(u,v)$---where the curve is not singular---the
branch points are isolated, or, equivalently, $\del_x f|_{x=r}
\neq 0$.  At these points a good complex coordinate on the curve
is $\xi=\sqrt{x-r}$, so that $dx = 2\xi d\xi + \del_u r du + \del_v r
dv$, and $y \sim \xi$.  
Plugging into (\ref{Omega1}) gives the pole part of the 2 form, 
$\Omega \sim (\del_u r - r\del_v r  + b) \xi^{-1} du \wedge dv$, which 
cancels when $\del_u r - r\del_v r  + b |_{x=r}=0$.  Since $r(u,v)$ 
is defined as a solution of $f|_{x=r}=0$, it follows that $\del_u r 
= -(\del_u f/\del_x f)|_{x=r}$ and $\del_v r = -(\del_v f / \del_x 
f)|_{x=r}$, so the condition for the pole to cancel can be written
\be\label{Omega2}
\left. \left( \del_u f - x\del_v f - b \del_x f
\right)\right|_{x=r}=0,
\ee
where we have used that $\del_x f|_{x=r}\neq0$ for generic $(u,v)$.
The integrability condition (\ref{rski}) evaluated at $x=r$
implies precisely (\ref{Omega2}), since $f|_{x=r}=0$, and thus
ensures the holomorphy of $\Omega$.

\subsection{Scaling limit}

To make further progress in solving the equations (\ref{rski}) and 
(\ref{effexp}), we now scale 
towards the singularity.  By constraining the $u$ and $v$ dependence 
of the curve, and by fixing all but a 3-parameter subgroup of the 
holomorphic reparametrizations, this scaling dramatically simplifies 
the BPS integrability equation (\ref{rski}).

Isolate the $y^2\sim x^6$ singularity by scaling $u\sim Z^{[u]}$, 
$v\sim Z^{[v]}$, and $x\sim Z^{[x]}$, as $Z\to0$.  Since the central 
charge $Z$ has dimensions of mass, $[u]$ and $[v]$ are the usual 
mass dimensions.  We are taking $Z\to0$ since BPS masses must vanish 
at the singularity.  Furthermore, since the singularity has been fixed 
at $u=v=x=0$, these scaling dimensions must all have positive real
parts,
\be\label{posre}
\Re[u],\ \Re[v],\ \Re[x]>0,
\ee
so that $Z\to0$ scales to the singularity.  Equations (\ref{crv6}) and 
(\ref{rsk}) imply the following relations among scaling dimensions:
\be\label{dim}
\begin{array}{rlrl}
{} [a] =& -2[v]+4[u]-2,&\qquad\qquad
{} [b] =& [v]-2[u],\\
{} [f] =& 6[v]-6[u],&\qquad\qquad
{} [f_k] =& (6-k)([v]-[u]),\\
{} [x] =& [v]-[u],&\qquad\qquad
{} [y] =& 2[v]-[u]-1.
\end{array}
\ee
It will be useful to define the following combinations of the $u$ 
and $v$ scaling dimensions:
\be\label{def}
r := [v]/[u] ,\qquad s :=1/[v].
\ee

Scaling, together with regularity, greatly constrain 
the form of the $f_k$.  Since each $f_k$ in (\ref{crv}) is regular in 
$u,v$, and vanishes at the origin, its terms must be of the form 
$u^n v^m$ for $n,m\in\Z\ge0$ with at least one of $n$ or $m$ not
equal to zero.  Comparing the scaling 
dimension of $u^n v^m$ with that of $f_k$ in (\ref{dim}) implies 
\be\label{rdisc}
n+mr=(6-k)(r-1).
\ee
In particular, we see that $r$ can take only discrete, rational values.
This further implies that $f_k$ is at most a $(5-k)$th 
order polynomial in $v$; more explicitly, this says that the $f_k$'s 
can only have the forms
\bea\label{fexp}
f_5 &=& f_{50}u^{r-1},
\nonumber\\
f_4 &=& f_{40}u^{2r-2} + f_{41}vu^{r-2},
\nonumber\\
f_3 &=& f_{30}u^{3r-3} + f_{31}vu^{2r-3} + f_{32}v^2u^{r-3},
\\
f_2 &=&  f_{20}u^{4r-4} + f_{21}vu^{3r-4} + f_{22}v^2u^{2r-4} 
+ f_{23}v^3u^{r-4},
\nonumber\\
f_1 &=& f_{10}u^{5r-5} + f_{11}vu^{4r-5} + f_{12}v^2u^{3r-5} 
+ f_{13}v^3u^{2r-5} + f_{14}v^4u^{r-5},
\nonumber\\
f_0 &=& f_{00}u^{6r-6} + f_{01}vu^{5r-6} + f_{02}v^2u^{4r-6} 
+ f_{03}v^3u^{3r-6} + f_{04}v^4u^{2r-6} + f_{05}v^5u^{r-6}.
\nonumber
\eea
Thus the scaling condition has further tightened the possible $u$, $v$
dependence of $f$ to be not just regular at the origin, but actually 
polynomial of order at most 5 in $v$.  Furthermore, since only 
non-negative integer $u$ exponents are allowed by the single-valuedness
of the $f_k$, we learn that $r$ must be real.  Then, since $[v]=[x]+[u]$,
(\ref{posre}) implies that actually $r>1$.  Thus only those terms in 
(\ref{fexp}) with positive integral exponents for $u$ and $r>1$ are 
allowed.

With this result, we have reduced our problem of finding all $f_k$'s 
satisfying the integrability condition (\ref{rski}) to an algebraic 
problem.  For, as we saw in the discussion after equation (\ref{effexp}), 
we can algebraically eliminate $a(u,v)$ and $b(u,v,x)$ in favor of 
$f(u,v,x)$.  Then by expanding the resulting differential equations 
for $f$ out in powers of $x$ and $v$, we get a set of algebraic 
equations for the coefficients of the $f_k$ polynomials.  The details
follow:

The scaling dimensions (\ref{dim}) motivate the following definitions 
of scale-invariant quantities $\a$, $\b$, $\phi$, $\o$, $\x$, and $\eta$, 
corresponding to $a$, $b$, $f$, $v$, $x$, and $y$, respectively:
\be\label{sca}
\begin{array}{rlrlrl}
a :=& u^{4-2r-2rs}\a,&\qquad
b :=& u^{r-2}(r\b+(1-r)\x),&\qquad
f :=& u^{6r-6} \phi,\\
v :=& -{1\over r} u^r \o,&
x :=& u^{r-1} \x,&
y :=& u^{2r-1-rs}\eta = u^{2r-1-rs}\sqrt{\a\phi}.
\end{array}
\ee
Note that $\a$ is a function of $\o$ only, and $\b$ and $\phi$
are functions of both $\o$ and $\x$.  Furthermore, since $b$ and 
$f$ are polynomials in $x$, $\b$ and $\phi$ must be polynomials 
in $\x$ of the same order.  We write
\be\label{phiexp}
\phi =  \x^6 + \sum_{k=0}^5 \x^k \phi_k,
\qquad\qquad\qquad
\b = \sum_{a=0}^3 \x^a \b_a.
\ee
The $\phi_k$ are the polynomials in $\o$, given by setting $u\to1$
and $v\to-\o/r$ in (\ref{fexp}).
Change to the scaling variables, $\{u,v,x\}\to \{u,\o,\x\}$, so
$\del_u \to \del_u -r (\o/u)\del_\o + (1-r) (\x/u)\del_\x$,
$\del_v \to -(r/u^r) \del_\o$, $\del_x \to u^{1-r} \del_\x$,
and 
the integrability condition (\ref{rsk}) becomes
\be\label{rsk2}
\left[ (\x-\o) \del_\o +(s-1)\right] \eta^{-1}
=  \del_\x \left(\b \eta^{-1}\right) .
\ee
It is remarkable that only the dimension of $v$ (through $s$)
enters this equation, and that $[u]$ does not.
Substituting $\eta^2=\a\phi$ gives
\be\label{rsk1}
(\x-\o)\del_\o\phi=\b \del_\x\phi
-\phi[2(1-s)+2\del_\x\b+(\x-\o)\del_\o\ln\a].
\ee
Expand $\phi$ and $\b$ as in (\ref{phiexp}).
The coefficients of each power of $\x$ should then
vanish separately, giving 9 potential equations in $\o$
corresponding to $\x^n$ for $n=0,1,\ldots,8$ since we
have seen that $\phi$ is a 6th order and
$\b$ is a 3rd order polynomial in $\x$.
The $\x^8$ equation is identically satisfied.
The $\x^7$ equation gives
\be\label{whataa}
\del_\o\ln\a = 2\b_2-\phi_5\b_3,
\ee
where $\phi_5$ is the coefficient of $\x^5$ in 
$\phi$.
This allows us to eliminate $\a$ from (\ref{rsk1}),
giving an equation for $\phi$ and $\b$ alone.
Moreover, since no $\del_\o$ derivatives of $\b$ 
enter, the $\beta$ can be eliminated algebraically,
leaving a system of polynomial equations in $\o$.

\subsection{Reparametrization invariance, revisited}

Before we reduce (\ref{rsk1}) to polynomial equations, though, 
we should fix the remaining holomorphic reparametrization 
symmetries (\ref{reparam}) of our system.  These are largely 
fixed by imposing scaling.  In fact, the only redefinitions of 
$u$ and $v$ preserving their scaling dimensions and regular at 
$u=v=0$ are (since $r>1$)
\be\label{holrep}
u \to A u,\qquad v \to A^r (B v + C u^r) .
\ee
(We put the factor of $A^r$ in the redefinition of $v$ for later 
convenience.)
Furthermore, the shift in $v$ with coefficient $C$ is only allowed 
when $r\in\Z$ by regularity.  This redefinition induces by 
(\ref{reparam}) 
\be\label{holrep2}
x \to A^{r-1} (B x -r C u^{r-1}),
\qquad
y \to A^{2r-1} B^2 y .
\ee

Consider just the $C$ shift (setting $A=B=1$ for the moment).  The 
shift in $v$ just reorganizes the coefficients within the $f_k$, while 
the shift in $x$ shifts the coefficients of the $f_k$ by multiples 
of the coefficients of the $f_j$ with $j>k$, as well as a constant 
shift from the $x^6$ term.  In particular, by choosing $C=f_{50}/(6r)$, 
we can shift away $f_5$ altogether.  Since, by (\ref{fexp}), $f_5$ was 
only non-vanishing for $r\in\Z$ anyway, we can fix the $C$ shift by the 
gauge choice
\be\label{Cgauge}
f_5(u,v)=0.
\ee

The other two reparametrizations, with $A,B\neq0$ and $C=0$, give rise
to non-trivial rescalings of the curve.  In particular, it is not too 
hard to check that when $B=1$, the rescaling does not change 
$f(x,u,v)$ in $y$, but only changes the prefactor $a(u,v) \to A^{2r-4} 
a(Au, A^r v)$.  Since $a$ has dimension $4[u]-2[v]-2$ by (\ref{dim}), 
this just amounts to an overall rescaling of $a\to A^{-2rs}a$.  Since 
$rs\neq0$, this can be used to set the overall normalization of $a$ to 
any desired value; this is equivalent to fixing the integration constant 
in the $\del_{u,v}\ln a$ equations.  The remaining reparametrization, 
found by, say, setting $A=1$ and $B\neq1$, changes $f$ by
\be\label{Bgauge}
f(x,u,v) \to B^{-6} f(Bx,u,Bv) .
\ee
Since by (\ref{fexp}) $f$ is not homogeneous in $x$ and $v$, this freedom
can be used to set one coefficient in $f$ besides $x^6$ to 1.

This 3-parameter group of holomorphic reparametrizations 
is easily identified as a symmetry group of the scale-invariant
integrability equation (\ref{rsk2}):
if $\{\eta(\x,\o),\b(\x,\o)\}$ is any solution, then so is 
$\{ A\eta(B\x+C , B\o+C) , B^{-1}\b(B\x+C,B\o+C) \}$ for 
complex $A$, $B$, $C$. 

So, fixing the shift symmetry by the choice (\ref{Cgauge}) sets
$\phi_5=0$.  Then eliminating $\del_\o\ln\a$ from (\ref{rsk1})
using (\ref{whataa}) gives
\be\label{rsk1.5}
(\x-\o)\del_\o\phi=\b \del_\x\phi
-2\phi[(1-s)+\del_\x\b+(\x-\o)\b_2].
\ee
The $\b_a$'s can be eliminated from this equation, giving a
set of 33 over-determined polynomial equations for the 20 remaining
coefficients of $\phi$, as well as $s$; see appendix A for details
on the algebraic elimination of the $\b$'s.  

On the one hand, this is 
good  news:  the system is highly over-constrained, and so we would 
expect it to have a relatively few very special solutions.  The bad 
news is that the equations are all 3rd to 5th order polynomial 
equations in the unknowns, and rather algebraically opaque.  
We haven't been able to find the general solution of these equations;
we discuss the problem further in appendix A.  We can, nevertheless,
find many solutions by making certain simple guesses for the form of
$\b$.  These guesses, described in appendix B, give rise to
47 curves which solve the integrability equation.  16 of these
are clearly unphysical, for reasons discussed there.  The remaining 
31 potentially
physical curves are listed in table \ref{tab.crv}, also in appendix B.  
Note that to each curve there corresponds an infinite series of distinct 
scale invariant RSK geometries, labeled by their possible values of $[u]$. 
This is because the integrability equation (\ref{rsk1.5}) depends only
on $[v]$ (through the parameter $s$) but not on $[u]$; the regularity
condition on the curve, discussed above, restricts $r$, and therefore
$[u]$, to take discrete values (\ref{rdisc}).


\section{Further consistency conditions\label{s4}}

So far we have only demanded integrability of the BPS
equation (\ref{bps}) for the central charge $Z$.  Now that
we have a list of curves satisfying the integrability equation,
we can integrate to find $Z$, and examine its behavior in
the vicinity of the origin of $\MM$.  There have been many
studies of BPS spectra near singularities on the Coulomb branch, 
some of which evaluate elaborate consistency conditions 
involving curves of marginal stability and global monodromies on 
the Coulomb branch \cite{bpsspec}.  We will, instead, impose a much 
simpler and more basic consistency condition.  Using this condition,
we will show that only a finite number of curves (namely, 12) out of the 31 
infinite families of potentially physical RSK geometries can have a 
consistent physical interpretation as low energy effective actions.

This physical consistency condition is simply that for each 
singularity in the curve there must exist corresponding massless 
BPS states.  Complex degenerations (singularities) of the curve 
translate, by (\ref{varper}), to divergences in the $\U(1)^2$ 
gauge couplings in the effective theory on the Coulomb branch.  
On the other hand, such divergences in an effective action mean 
that extra charged states are becoming massless at the singularity.  
However, the BPS masses, computed by the central charge $Z$, give a 
lower bound on the masses of all charged states.  Therefore, for a 
consistent interpretation as an effective action on the Coulomb
branch, at each singularity of the curve we must have vanishing
central charges for certain charge sectors.  We refer to
this as the $Z$-consistency condition in what follows.

We will see that the vast majority of our scale-invariant RSK 
geometries do not sastisfy the $Z$-consistency condition: they 
lack massless BPS states 
required to explain the existence of certain one-dimensional curves 
of singularities in $\MM$ in the vicinity of the origin.
The way the $Z$-consistency condition just manages to leave
a finite number of solutions hinges on a delicate cancellation
between the behaviors of the curve prefactor $\a(\o)$ and the
body of the curve, $\phi(\x,\o)$.  This cancellation works because
of an identity satisfied by $[v]$ which is not at all obvious from
the derivation of the curves.

There are two other singular kinds of behavior of our curves
which one might consider unphysical, or perhaps novel behaviors
which could be physically interesting in their own right.  One possible 
strange behavior is that BPS masses might diverge as one approaches the 
singularity at $u=v=0$, but only from certain directions.  Another is that
the distance to the singularity at $u=v=0$ might diverge in 
the metric induced on $\MM$.  This would indicate that the ``point" 
$u=v=0$ does not have 
an interpretation as a vacuum of some $N=2$ theory, but would rather 
represent another asymptotic limit of the Coulomb branch.
But, in fact, the following analysis will show that neither of
these strange behaviors occur for any of our curves.

\subsection{Central charge}

So far we have solved the local integrability equation for the
existence of the BPS masses (central charge).  Here we 
integrate the BPS equation (\ref{bps}) and examine the behavior
of the masses near the singularity.

By scaling, the central charge can be written $Z = u^{rs} \zeta(\o)$.
Inserting this in the BPS equations (\ref{bps}) gives 
\be\label{intbps}
\zeta  = {1\over rs} \oint { (\x-\o) d \x \over \sqrt{\a\phi}},
\qquad
\del_\o \zeta = -{1\over r}\oint {  d\x \over \sqrt{\a\phi}}.
\ee
The compatibility between the first and second equations
is precisely our integrability equation in the form (\ref{rsk2}).
The first equation in (\ref{intbps}) implies that masses are 
given by
\be\label{cc}
Z = {u^{rs}\over rs} {1\over\sqrt{\a(\o)}}
\oint  { (\x-\o) d \x \over \sqrt{\phi(\x,\o)}} .
\ee

We are interested in the behavior of the BPS masses in the
vicinity of the singularity at $u=v=0$.  This point can be
approached from different directions in the $u$-$v$ plane
by scaling $u\to0$ with $\o=\o_0$, where $\o_0$
is any fixed value in the $\o$-plane.  For generic $\o_0$,
$\a$ and the integrand are finite, so $Z\sim u^{rs}\to0$.
(Recall that the scaling assumption implies $\Re[u]>0$,
and so, since $rs=1/[u]$, $\Re(rs)>0$ as well.)  

\TABLE[ht]
{\begin{tabular}{||l|l|c|l||l|l|c|l||} \hline
Curve	&Discriminant&\multicolumn{2}{l||}{$Z$-consistent at}	&
Curve	&Discriminant&\multicolumn{2}{l||}{$Z$-consistent at}	\\ 
$[v]=$			&$\Delta[\phi]\propto$	&$v=0$&$u=0$				&
$[v]=$			&$\Delta[\phi]\propto$	&$v=0$&$u=0$		\\ \hline\hline
\sm$1$			&\sm1					&$\surd$	&\sm$r=1$		&
\sm$3\ \ \,b,b'$	&\sm$\o^{20}$			&		&\sm$r=3$		\\ \hline
\sm$6/5$			&\sm$\o^5$				&		&\sm$r={6/5}$	&
\sm$10/3$		&\sm$\o^{16}(\o^2+1)$		&$\surd$	&\sm$r={5/2}$	\\ \hline
\sm$5/4$			&\sm$\o^6$				&		&\sm$r={5/4}$	&
\sm$15/4$		&\sm$\o^{18}(\o^2+1)$		&		&\sm$r=3$		\\ \hline
\sm$3/2\ \ \ \,a$&\sm$(\o^5+1)$			&$\surd$	&\sm$r={6/5}$	&
\sm$4$			&\sm$\o^{12}(\o^3+1)$		&$\surd$	&\sm$r=2$		\\ \hline
\sm$3/2\ b,b'$	&\sm$\o^{10}$			&		&\sm$r={3/2}$	&
\sm$5\ \ \ \ \ a$&\sm$\o^9(\o^3+4)$		&$\surd$	&\sm$r={5/3}$	\\ \hline
\sm$5/3\ \ \ \,a$&\sm$\o^2(\o^4+1)$		&$\surd$	&\sm$r={5/4}$	& 
\sm$5\ \ \ \ \ b$&\sm$\o^{24}$			&		&\sm$r=5$		\\ \hline
\sm$5/3\ \ \ \,b$&\sm$\o^{12}$			&		&\sm$r={5/3}$	&
\sm$6\ \ a,a'$	&\sm$\o^{18}(\o^2+\t\o+1)$&$\surd$&\sm$r=3$		\\ \hline
\sm$15/8$		&\sm$\o^6(\o^4+1)$		&		&\sm$r={3/2}$	&
\sm$6\ \ \ \ \ b$&\sm$\o^{25}$			&		&\sm$r=6$		\\ \hline
\sm$2,\ 2'$		&\sm$\o^{15}$			&		&\sm$r=2$		&
\sm$20/3$		&\sm$\o^{23}(\o+1)$		&$\surd$	&\sm$r=5$		\\ \hline
\sm$20/9$		&\sm$\o^9(\o^3+1)$		&$\surd$	&\sm$r={5/3}$	&
\sm$15/2\ a$		&\sm$\o^{16}(\o^2+4)$		&$\surd$	&\sm$r={5/2}$	\\ \hline
\sm$5/2\ \ \ \,a$&\sm$\o^{12}(\o^3+1)$	&		&\sm$r=2$		&
\sm$15/2\ b$		&\sm$\o^{24}(\o+1)$		&		&\sm$r=6$		\\ \hline
\sm$5/2\ \ \ \,b$&\sm$\o^{18}$			&		&\sm$r={5/2}$	&
\sm$12$			&\sm$\o^{24}(\o+1)$		&$\surd$	&\sm$r=6$		\\ \hline
\sm$3\ \ \ a,a'$	&\sm$\o^6(\o^4+\t\o^2+1)$	&$\surd$	&\sm$r={3/2}$	&
\sm$15$			&\sm$\o^{23}(\o+4)$		&$\surd$	&\sm$r=5$		\\ \hline
\end{tabular}
\caption{Discriminants of the curves as functions of the scaling
variable $\o$ on $\MM$.  Curves are labelled by $[v]$ and ``$a$",
``$b$", \etc, as in table 4, appendix B.  Curves with appropriate 
massless BPS states along any $v=0$ singularity are checked.  The 
values of $r$ for which curves have no singularity at $u=0$ are also 
shown.\label{tab.sing}}}

But there are special values of $\o_0$ where the behavior of $Z$ is 
more delicate: if $\o_0$ is a zero of $\a(\o)$ or is a point where 
two or more zeros of $\phi(\x,\o)$ coincide, then, potentially, $Z$ 
could remain finite or even diverge.   
Note that the prefactor, $\a$, of the curve has no effect on the 
effective gauge couplings since it cancels out in (\ref{tdef1}). 
Thus complex degenerations of the curve are determined solely
by $\phi$.   To locate for which $\o$ the zeros of $\phi$ collide, 
we compute  the discriminant of $\phi(\x)$ as a function of $\o$ (dropping
overall factors which might depend on the $\t_i$), shown in table
\ref{tab.sing}.  The discriminant vanishes whenever two or more 
roots of $\phi$ coincide in the $\x$-plane.  

\paragraph{Singularities along $\bf v=0$.}

All of the curves in table \ref{tab.crv} have the general form
\be\label{leading}
\phi \sim \x^6 + c\, \o^j\x^k + \ldots + c'\,\o^\ell\x^m,
\ee
for some unique $5 \ge \ell\ge j\ge0$, where $k$ is the largest power 
of $\x$ appearing (besides the leading $\x^6$ term), and $m$ the 
smallest.  (This includes the cases where the $\o^j\x^k$ and $\o^\ell\x^m$ 
terms are the same.)  This is mostly just a definition of $j$, $k$, 
$\ell$, and $m$; the only nontrivial facts---those following from a 
case-by-case examination of the curves in table \ref{tab.crv}---are 
the uniqueness and ordering of $j$ and $\ell$.  Indeed, every curve 
has just a single power of $\o^p$ for each $\x^q$, and $p$ increases 
as $q$ decreases.  Another non-trivial fact, noted in appendix B, is 
that the curve prefactor $\a$ is related to the $j,k$ of (\ref{leading})
by
\be\label{alphaomega}
\a^{-1/2}=\o^{j/(6-k)}.
\ee

{}From table \ref{tab.sing} we see that, except for the $[v]=1$ and 
$3/2a$ curves, all curves have colliding roots at $\o=0$.  This 
corresponds to a 1-complex-dimensional line of singularities in the 
Coulomb branch along $v=0$ (thus including the singular point at
$u=v=0$).  In particular, we see from (\ref{leading}) that as 
$\o\to0$, $6-k$ roots of $\phi$ collide at $\x=0$, scaling with 
$\o$ as $\x\sim\o^{j/(6-k)}$.
Now evaluate $Z$ for any contour looping around these colliding zeros 
by making the change of variables $\x=\o^{j/(6-k)} x$ in (\ref{cc}):
\be\label{Zcancel}
Z \sim u^{rs} \o^{j/(6-k)}
\oint  { (\o^{j/(6-k)} x-\o) \o^{j/(6-k)} dx 
\over \o^{3j/(6-k)}\sqrt{x^6+c x^k +\ldots}} ,
\ee
where we have used (\ref{alphaomega}).  Taking the $\o\to0$ limit then 
gives a finite answer for $Z$ (at non-zero $u$).  Thus, the $\a$ prefactor 
precisely cancels the potential divergence coming from the leading 
singularity in $\phi$ at $\o=0$.

However, the precise cancellation in (\ref{Zcancel}) means that 
at $u\neq0$ and $v=0$, there is a singularity in the complex structure 
of the curve, but no massless charged states, thus failing the Z-consistency
test.  This conclusion does not
hold, though, when there are subleading terms in $\phi$ beyond the
$\o^j\x^k$ term in (\ref{leading}).  For when there are, some of the 
other $k$ roots of $\phi$ scale to zero as a different power of $\o$. 
In the case $k=1$, then the scaling in (\ref{Zcancel}) still holds for
all possible vanishing cycles, and the Z consistency test fails.  However,
when $k\ge2$, there are two or more roots which collide more quickly
as $\o\to0$ and so $Z$ evaluated for contours encircling these roots 
then do, in fact, vanish in this limit.  The curves which do not
fail the $Z$-consistency condition at $v=0$ are marked with a $\surd$ 
in table \ref{tab.sing}.  

\paragraph{Singularities along $\bf v\sim u^r$.}

Many of the curves also degenerate at values of $\o\neq0$, as can be 
seen from their discriminants in table \ref{tab.sing}.  Recalling
that $\o\sim v/u^r$, such degenerations give singularities in
the Coulomb branch along 1-complex-dimensional lines of the form
$v\sim u^r$.  All these degenerations are due to a pair of roots of 
$\phi$ colliding, say at $\x=\x_0$, as $\o\to\o_0$.  A case-by-case
check shows that $\x_0=\o_0$ for these curves.  Changing variables 
in (\ref{cc}) to $\o=\o_0+\d$, $\x=\o_0+\d^{1/2} x$ and taking the 
limit as $\d\to0$, it follows that $Z\sim \d^{1/2}\to0$ for a contour 
encircling the two colliding roots.  So these degenerations always 
lead to massless states, and never to massive ones.

\paragraph{Singularities along $\bf u=0$.}

Finally, we must also examine the $\o\to\infty$ limit of $Z$,
which can correspond either to approaching $u=0$ with $v$ fixed
or $v=\infty$ with $u$ fixed.  It is only the former limit that
we are interested in.  
As $\o\to\infty$, for all except the $[v]=1$ curve, 6 or 5 roots of 
$\phi$ approach $\x=\infty$.  
In all cases it is the lowest power of $\x$ in $\phi$ which controls 
the scaling of the roots to $\x=\infty$.  (This is not obvious 
{\em a priori}, but amounts to the statement $\ell/(6-m) \ge j/(6-k)$,
and is checked to hold for each of our curves.)  Then the roots going 
to infinity scale as $\x \sim \o^{\ell/(6-m)}$.  Plugging these into 
(\ref{cc}) then gives
\be\label{Zcancel2}
Z \sim u^{rs}\o^{j/(6-k)+1-2\ell/(6-m)}
\sim u^{r[s+2\ell/(6-m)-1-j/(6-k)]},
\ee
where in the second step we used that $\o\sim u^{-r}$ at fixed $v$.
Now, another fact about the curves in table \ref{tab.crv}, checked 
for each case, is that
\be\label{Zcancel3}
s:={1\over[v]} = 1 + {j\over 6-k} - {2\ell\over 6-m} .
\ee
This remarkable identity, not at all obvious from our derivation of the 
curves, implies that the exponent in (\ref{Zcancel2}) vanishes, and so 
these BPS masses remain finite, apparently failing the $Z$-consistency 
condition. 

There is a loophole, however:  we have shown that there are no
massless charged states, but we haven't actually shown that there
is a singularity in the curve at $u=0$ for finite $v$ on the Coulomb 
branch.  Such a singularity does not follow from that fact that
roots of $\phi$ collide at $\o\to\infty$ since, as we mentioned
above, $\o=\infty$ corresponds either to a $u=0$ line or to
an uninteresting $v=\infty$ limit on the Coulomb branch.  To tell
which is realized, we must put the $u$-dependence back into our
curves.  This is easily done by comparing to (\ref{fexp}) and
(\ref{sca}): the $\o^\ell\x^m$ term which controls the $\o\to\infty$ 
behavior becomes $u^{(6-m-\ell)r-(6-m)} v^\ell x^m$ once factors of
$u$ are restored.  Precisely for
\be\label{Zcancel4}
r={6-m\over 6-m-\ell}
\ee
the $u$-dependence drops out, and no roots of the curve collide
as $u\to0$ at fixed $v$.  So it is only for these values of $r$,
listed in table \ref{tab.sing}, that the curves pass the $Z$-consistency
test at $u=0$.  

The net result of this analysis is that only a finite number of 
solutions of the integrability equation also satisfy the
$Z$-consistency condition.  These are the curves with a $\surd$
in table \ref{tab.sing} and which have a representative for
the listed value of $r$.  Only the $[v]=1$, $3/2b'$, $2'$, $3a'$,
$3b'$, and $6a'$ curves do not satisfy the latter requirement; in
all other cases, the listed $r$ is the minimum value of $r$ allowed
in table \ref{tab.crv}.  The resulting 12 curves satisfying the
$Z$-consistency condition are listed, along
with their scaling dimensions $[u]$ and $[v]$, below in table 
\ref{tab.cft}.  There we have converted them from the scaling variables
back to the original $x,y,u,v$ ones.  We have also taken the liberty
to make a some $A$ and $B$ rescalings (\ref{holrep}) of $u$ and
$v$ in order to simplify some coefficients in the curves.

\paragraph{Comments.}

The $Z$-consistency condition gives no restriction on the allowed
curves in the classification
of scale-invariant 1-dimensional RSK geometries, reviewed in section
\ref{s2}.  The reason is simply that complex singularities in a 
1-dimensional Coulomb branch are isolated, while the singularity
at the origin of a 2-dimensional Coulomb branch is at the intersection
of 1-dimensional lines of singularities.  The $Z$-consistency condition
concerns the existence of appropriate massless BPS states at these
lines of singularities, and so is trivially satisfied in the
1-dimensional case.

Finally, it is worth pointing out that the general algebraic form of 
the integrability equation permits much more singular behavior of the 
prefactor $\a$.  In particular, (\ref{whataa}) implies that $\a$ 
is determined by a relation of the general form $\del_\o\ln \a = 
\til R(\o)$ for some rational function $\til R$ (ratio of polynomials)
in $\o$.  This follows because the $\b_a$ are given as ratios 
of polynomials; see the discussion of the elimination of the $\b_a$ 
in appendix A.  Integrating such an equation gives $\a(\o) = e^{R(\o)}
\prod_i (\o-\o_i)^{a_i}$ for some other rational $R$ and some $\o_i$ 
and $a_i$.  Now, any nontrivial $e^{R(\o)}$ dependence in $\a$ would 
lead to diverging BPS masses from some direction of approach to the 
origin, since every nontrivial $R$ has a pole at some $\o_0$ or at 
infinity, where $\Re(e^R)\to+\infty$, dominating the behavior of $Z$.
It is striking that none of the solutions of the integrability equation 
have this more generic, unphysical behavior.

\subsection{Metric}

It is easy to examine at this point the behavior of the metric on
the Coulomb branch in the vicinity of the $u=v=0$ singularity.  A
physical argument predicts how the metric should behave.  The metric
is induced by the kinetic term for the massless neutral scalars whose
vevs are $u$ and $v$.  Since the fluctuations of free scalar fields
scale as a mass (\ie, have mass-dimension 1), geodesic lengths should
be assigned scaling dimension 1.  So scaling near the origin
of $\MM$ implies that along geodesics out of the origin, $u\sim t^{[u]}$ 
and $v\sim t^{[v]}$, where $t$ is the geodesic length.  Since $\Re
[u],\Re [v] >0$, this implies finite distance to the singularity and 
only mild (conical) singularities.

These predictions are easy to confirm explicitly, and are insensitive
to the details of the curves.
The metric on $\MM$ is induced from the kinetic term for the
scalar fields.  In special coordinates, $a^i$, these terms are
$\Im(\t_{ij} \del_\mu a^i \del^\mu \bar a^\jb)$.  In the general, 
$u^i$, coordinates this becomes $\Im[\t_{ij} (\del a^i/\del u^k)
(\del \bar a^\jb/\del \bar u^{\bar m}) \del_\mu u^k \del^\mu \bar 
u^{\bar m}]$.  Thus the metric on $\MM$ is $ds^2=g_{i\jb}
du^i d\bar u^{\jb}$, with
\be
g_{i\bar\jmath} = {1\over2i}\sum_k \left(
\oint_{\a^k}\bar{\o_j} \oint_{\b_k} \o_i
-\oint_{\a^k}\o_i \oint_{\b_k} \bar{\o_j}\right)
= {i\over2} \int_T \o_i\wedge\bar{\o_j}\wedge t^{r-1}
= {i\over2} \int_\Sigma \o_i\wedge\bar{\o_j},
\ee
where we have used (\ref{tdef1}) and (\ref{BPS}) in the first equality, 
and translated it to an integral over the fibered torus $T$ or the 
Seiberg-Witten Riemann surface $\Sigma$ in the second or third equality.
In the 2-dimensional case we are examining, this translates to
\be
\pmatrix{g_{u\bar u} & g_{u\bar v} \cr g_{v\bar u} & g_{v\bar v}\cr}
= {i\over2}\int_\Sigma {dx\wedge d\bar x\over y\bar y}
\pmatrix{x\bar x & x\cr \bar x & 1\cr} = \pmatrix{
u^{rs-1} \bar u^{rs-1} \a(\o,\bar\o) & 
u^{rs-1} \bar u^{rs-r} \b(\o,\bar\o) \cr 
u^{rs-r} \bar u^{rs-1} \bar\b(\o,\bar\o) & 
u^{rs-r} \bar u^{rs-r} \g(\o,\bar\o) \cr},
\ee
where in the second equality we have changed to the scaling
variables; $\a$, $\b$, and $\g$ are scale invariant functions
whose specific form is irrelevant since they will cancel in
the end.  

The geodesic equation is $0 = g_{\ib j} \udd^j + \del_j g_{\ib k} 
\ud^j \ud^k$, where the dot represents a derivative with respect to 
the length $t$.  This is easy to integrate given the scaling form 
of the curves.  Changing variables from $(u, v)$ to $(u, \o)$,
and guessing that the geodesic is $u=u(t)$ with $\o$ constant, 
the $i=u$ component of the geodesic equation becomes $0 = (\a - \o 
\bar\b) \left[ \udd u + (rs-1) \ud \ud \right]$ (and a similar 
equation for the $i=v$ one).  The solution to this equation is
$u \propto t^{1/(rs)} = t^{[u]}$ at fixed $\o$, which is precisely
the result predicted by physical reasoning.

\section{Discussion\label{s5}}

We can use our result, table \ref{tab.cft}, to test the hypothesis
that all consistent scale-invariant RSK geometries describe the
Coulomb branch effective actions of superconformal field theories (SCFTs).
Are all the curves in table \ref{tab.cft} consistent with conformal 
invariance (as opposed to scale invariance, which is what we imposed)?
Do known $N=2$ SCFTs appear on our list?
The answers we find are that all our scale-invariant RSK geometries
are consistent with an interpretation as effective theories of
$N=2$ superconformal field theories, and, where we can check, they 
exist as quantum field theories.  We then go on to discuss directions
for further research.

\TABLE[ht]
{\begin{tabular}{||ll|rl||} \hline
\multicolumn{2}{||l|}{Dimensions}&\multicolumn{2}{l||}{Curve}	\\
$[u]$&$[v]$	&\multicolumn{2}{l||}{$y^2= \ldots$}			\\ \hline\hline
5/4	&3/2 	&			&$[x^6+(ux+v)]$					\\ \hline 
4/3	&5/3		&			&$[x^6+x(ux+v)]$					\\ \hline 
4/3	&20/9	&$v^{-1/2}$	&$[x^6+vx(3ux+4v)]$				\\ \hline
2	&3		&			&$[x^6+\t x^3(ux+v)+(ux+v)^2]$	\\ \hline 
4/3	&10/3	&$v^{-1}$	&$[x^6+v^2x(ux+2v)]$				\\ \hline
2	&4		&$v^{-1/2}$	&$[x^6+v(3ux+4v)^2]$				\\ \hline
3	&5		&			&$[x^6+x(ux+v)^2]$				\\ \hline
2	&6		&$v^{-1}$	&$[x^6+\t vx^3(ux+2v)+v^2(ux+2v)^2]$\\ \hline 
4/3	&20/3	&$v^{-3/2}$	&$[x^6+v^3x(ux+4v)]$				\\ \hline
3	&15/2	&$v^{-2/3}$	&$[x^6+vx(2ux+3v)^2]$				\\ \hline 
2	&12		&$v^{-3/2}$	&$[x^6+v^3(ux+4v)^2]$				\\ \hline
3	&15		&$v^{-4/3}$	&$[x^6+v^2x(ux+3v)^2]$			\\ \hline
\end{tabular}
\caption{Solutions of the integrability equation and
$Z$-consistency condition. For all curves the basis of holomorphic 
one-forms is $\o_u=xdx/y$ and $\o_v=dx/y$.\label{tab.cft}}}

\paragraph{Scaling {\em versus} conformal invariance}

Conformal invariance plus unitarity in a field theory implies
that the scaling dimensions of all interacting fields must
be real and greater than one \cite{scft}.  The fact that all values of
$[u]$ and $[v]$ in table \ref{tab.cft} satisfy this is evidence
that they can be consistently interpreted as SCFTs.  Our search
for singular RSK geometries did not impose this requirement
in any way.  Indeed, as the discussion of the last section 
shows, the way in which it is met is quite complicated.

It is worth noting that a variety of other scaling behaviors 
which are thought to violate conformal invariance in field 
theories \cite{pol88} are allowed in our classification, but 
were not found to occur.  For example, we allow complex scaling 
dimensions, which can describe discretely self-similar solutions
(which could be interpreted in terms of limit cycles in the RG 
flow of a field theory); but only real scaling dimensions are found.  
Also, curves for which $[v]$ is allowed to vary over a continuum 
of values are not found (except for one unphysical example mentioned 
in appendix B).

Finally, $[u]$ and $[v]$ in table \ref{tab.cft} only take on
rational values, though this was not demanded before-hand.
(The physical monodromy conditions described in section \ref{s3}
give a simple explanation of why the ratio $r=[v]/[u]$ must be 
rational; however, our polynomial system following from the 
integrability equation implies {\em a priori} only that when 
$[v]$ is discrete, it will be an algebraic number.)  

\paragraph{Matching to known conformal field theories.}

Known $N=2$ SCFTs are those with either a weakly coupled lagrangian 
description, or found by taking special limits of lagrangian theories.
There is a short list of $N=2$ conformal field theories with weakly 
coupled lagrangian limits and two-dimensional Coulomb branches.  These 
are all the rank 2 gauge theories with massless hypermultiplets in 
appropriate representations.  The curves are known for the $\SU(2) 
\times \SU(2)$ \cite{w9703}, $\SU(3)$ \cite{aps9505,lll9805}, and $\SO(5) 
\simeq \Sp(4)$ \cite{as9509,asty9611,dls9612} theories, but the curve 
for the scale-invariant $G_2$ theory (with four massless $\bf 7$'s) 
has not appeared in the literature so far.  Of the known curves, only 
the $\SU(3)$ ones have a $y^2=x^6$ singularity.  The others turn out to 
all have $y^2=x^5$ or $y^2=(x^2-1)^3$ singularities.  These latter 
correspond to the $(5,1)$ and $(3,3)$ singularity types mentioned in 
section \ref{s3}.3, which will be discussed elsewhere.\footnote{Note 
that the singularity type depends on the choice of basis of one-forms.  
In the analysis in this paper we have chosen the standard basis 
(\ref{forms}).  This does not always correspond to what is used in 
the literature, in which case an appropriate $\GL(2,\C)$ change of 
basis (\ref{genbas}) is required.}  The $\SU(3)$ curve indeed appears 
in table \ref{tab.cft}: it is the curve with $[u]=2$, $[v]=3$, and a 
marginal coupling $\t$.

The other curve in table \ref{tab.cft} with a marginal coupling
has $[u]=2$ and $[v]=6$, which are the scaling dimensions one
expects for a scale-invariant $G_2$ gauge theory.  This follows
because at weak coupling the dimensions will be given by the 
classical dimensions of the basic gauge invariants made from the 
adjoint scalar, which are given by the exponents (plus 1) of the 
Lie algebra, and they are protected from varying as the coupling
changes.  Furthermore, it is easy to check that this curve has 
the expected pattern of singularities and associated monodromies on 
the Coulomb branch.  It would be interesting to deform this curve
with masses to compare to the various proposed curves \cite{G2}
(not all of which agree) for the asymptotically free $G_2$ theories
with up to three hypermultiplets in the $\bf 7$.

Other SCFTs have been found by tuning moduli in asymptotically free 
(AF) lagrangian field theories.  One finds in this way
partially scale-invariant RSK geometries corresponding to 
sub-leading singularities, as described in section \ref{s2}.3 above.  
By dropping the dependence on the directions in the moduli space 
corresponding to $\U(1)$ factors under which no charged states
become massless, we can isolate fully scale-invariant RSK geometries
of lower dimension.  These then correspond to some SCFT embedded
in the larger AF theory.  It is computationally very hard to search
for such sub-leading singularities in a systematic manner; however
a few series of such SCFTs were found in \cite{higherdim}.
Among these, we find two rank 2 SCFTs by tuning in $\SU(5)$ and 
$\SU(6)$ Yang-Mills, with dimensions (8/7,10/7) and (5/4,3/2),
respectively.  The first is a $y^2=x^5$ singularity, so not on our 
list, while the second is the first entry in table \ref{tab.cft}.

Another strategy for finding SCFTs is to tune only masses
in lagrangian theories in a search for maximal degenerations
of the curve (to avoid sub-leading singularities).  This was
followed in \cite{apsw9511} to find the all the dimension
one RSK geometries with $[u]<2$ in table \ref{tab2} within
the $\SU(2)$ theory with 4 fundamentals.  It would be natural
to try the same strategy starting with the $\SU(3)$ curve with
6 fundamentals, and the $G_2$ curve with 4 fundamentals.  However,
a straightforward search for maximal degenerations in a 
two-dimensional Coulomb branch is algebraically prohibitively
complicated to carry out systematically.
Finally, just as SCFTs realizing the $E_n$ series of 1-dimensional 
scale-invariant RSK geometries in table \ref{tab2} were found using
string constructions \cite{excft}, it would be interesting to see if 
these constructions could be generalized to 2-dimensional Coulomb 
branches.

\paragraph{Finding all solutions to the integrability equation.}

An idea for completing the classification of 2-dimensional scale-invariant
RSK geometries (aside from running Buchberger's algorithm on a larger 
computer---see the discussion in appendix A), is to present the
genus 2 Riemann surface in a different way.  

The representation of the Riemann surface in terms of a single degree 6 
polynomial in $\CP^2_{(3,1,1)}$ simplified the analysis by making the 
degenerations of the Riemann surface simple to locate algebraically.
However, because it is in terms of variables in weighted projective space 
it treats the $y$ and $x$ variables differently.  
A more symmetrical description of the curves might lead to significant
simplifications in the algebraic form of the integrability equations.
The fact that all genus 2 surfaces may be realized as intersections of 
quadratic surfaces in $\CP^3$ \cite{gh78} suggests one route to addressing
this problem.

\paragraph{Deformations away from scale invariance.}

A natural next step, given a classification of SCFTs, is to
examine their deformations by relevant operators.  A local
classification of all possible such deformations preserving
the conditions of RSK geometry should be possible along
the lines of the present paper ({\em cf.}\ section 17 of the
second paper in \cite{sw94}).  A classification of large
deformations, though, is likely to be much more complicated.
For instance, such an analysis (in the context of webs of $(p,q)$ 
strings and 7-branes in IIB string theory \cite{strjnc}) for 
one-dimensional RSK geometries gives quite elaborate results.

Note that it is by no means guaranteed, in any case, that deformations
of SCFTs will give all physical RSK geometries.  There are
examples of RG flows that have no UV fixed points, at least with 
local quantum field theory interpretations.
Among these are little string theories \cite{lst}, and perhaps
the UV ``walls" of the cascading models of \cite{ks0007}.
In the case of little string theories, one low energy implication
is the existence of compact Coulomb branches \cite{i9708,gk9802}.
This sort of global structure on the Coulomb branch cannot be seen 
in perturbations from essentially local scale-invariant geometries.

\paragraph{Extension to higher-dimensional RSK geometries.}

The general approach we have developed here can be applied to 
higher-dimensional RSK geometries as well.  
%
Since all principally polarized
abelian varieties are realized as the Jacobian tori of genus
3 Riemann surfaces, we can use a Riemann surface description as
we did in the 2-dimensional case.  The generic genus 3 Riemann
surface can be described by a quadric in $\CP^2$; but this misses
the hyperelliptic curves which form a co-dimension one subvariety
in the space of genus 3 Riemann surfaces, and which can be
realized as the zeros of a degree 8 polynomial in $\CP^2_{(4,1,1)}$.

Note that for genus 3 curves we have a 3-dimensional space of
holomorphic one-forms, and so need nine complex parameters to
specify a basis.  The group of holomorphic reparametrizations
in $\CP^2$ is $\SL(3,\C)$, an eight complex-dimensional group.
(It is $\SL$ instead of $\GL$ because of the homogeneity
condition on the projective coordinates.)  This is not enough to
fix the basis of holomorphic one-forms to a canonical (non-singular)
form.  This may lead to a qualitatively new class of singularities
involving a degenerating basis of one-forms.\footnote{The hyperelliptic
case just avoids this at genus 3 since the group of holomorphic 
reparametrizations is 9 dimensional: 4 from $(w,x)$ $\GL(2,\C)$
transformations, 1 from rescaling $y$, 5 from shifting $y$ by
a degree 4 polynomial in $(w,x)$, and $-1$ from the projective
homogeneity condition.}



\appendix
\section{Derivation and form of polynomial equations}

In this appendix we will give the algebraic details of how the
BPS integrability equation (\ref{rsk1.5}) is converted to
an over-determined set of polynomial equations.  Unfortunately, the
resulting set of 33 5th-order equations in 21 variables is too
lengthy to reproduce here.  We will only outline some of the
main intermediate steps.  The interested reader can obtain a
{\it Mathematica} notebook with the derivation of the
full set of equations by emailing the authors.

Let us first recall the scale invariant equation.  For notational 
simplicity the substitution $s = 1 - t$ will be used, so that 
(\ref{rsk1.5}) becomes
\be\label{rsk1.5A}
(\x-\o)\del_\o\phi=\b \del_\x\phi
-2\phi[t+\del_\x\b+(\x-\o)\b_2],
\ee
where
\be
\phi = \x^6 + \sum_{k=0}^4 \x^k \phi_k(\o), \qquad\qquad
\b = \sum_{a=0}^3 \x^a \b_a(\o) .
\ee
Substituting for $\phi$ and $\b$ in (\ref{rsk1.5A}) gives a 
sixth-order polynomial in $\x$, each of whose coefficients 
must vanish:
\be\label{eqnlist1}
\begin{array}{lrrrrrl}
\x^0 :& 0=& &(\phantom{6\phi_0}{}-2\o\phi_0)\b_2 
&\!{}+2\phi_0\b_1 &\!{}-\phantom{2}\phi_1 \b_0
&\!{}+ 2t\phi_0\qquad\,\,{}-\o\phi_0'
\\
\x^1 :& 0=& &(6\phi_0-2\o\phi_1)\b_2 
&\!{}+\phantom{2}\phi_1\b_1 &\!{}-2\phi_2\b_0 
&\!{}+2t\phi_1+\phi_0' -\o\phi_1'
\\
\x^2 :& 0=& 6\phi_0\b_3+{}\!&(5\phi_1-2\o\phi_2)\b_2 
& &\!{}-3\phi_3\b_0 
&\!{}+2t\phi_2+\phi_1' -\o\phi_2'
\\
\x^3 :& 0=& 5\phi_1\b_3+{}\!&(4\phi_2-2\o\phi_3)\b_2 
&\!{}-\phantom{2}\phi_3\b_1 &\!{}-4\phi_4\b_0 
&\!{}+2t\phi_3+\phi_2'-\o\phi_3'
\\ 
\x^4 :& 0=& 4\phi_2\b_3+{}\!&(3\phi_3-2\o\phi_4)\b_2 
&\!{}-2\phi_4\b_1 & 
&\!{}+2t\phi_4+\phi_3' -\o\phi_4'
\\
\x^5 :& 0=& 3\phi_3\b_3+{}\!&(2\phi_4\phantom{{}+2\o\phi_5})\b_2 
& &\!{}-6\phantom{\phi_4}\b_0 &\qquad\ \ \,\,{}+\phi_4'
\\
\x^6 :& 0=& 2\phi_4\b_3+{}\!&(\phantom{2\phi_4}{}-2\o\phantom{\phi_5})\b_2 
&\!{}-4\phantom{\phi_4}\b_1 & 
&\!{}+2t ,
\end{array}
\ee
where $\phi_k':=\del_\o\phi_k$.
We can solve the $\x^{5,6}$ equations for $\b_0$ and $\b_1$ without 
making any assumptions on relations between the $\phi_k$'s.  Substituting
gives
\be\label{eqnlist2}
\begin{array}{lll}
\x^0:\ 0=\b_2(\phantom{36\phi_0}-18\o\phi_0-2\phi_1\phi_4)&
\!{}+\b_3(\phantom{3}6\phi_0\phi_4 -\phantom{1}3\phi_1\phi_3) &
\!{}+18t\phi_0 \phantom{{}+6\phi_0'}-6\o\phi_0'-\phantom{1}\phi_1\phi_4'
\\
\x^1:\ 0=\b_2(36\phi_0-15\o\phi_1-4\phi_2\phi_4)&  
\!{}+\b_3(\phantom{3}3\phi_1\phi_4-\phantom{1}6\phi_2\phi_3)&
\!{}+15t\phi_1+6\phi_0' -6\o\phi_1'-2\phi_2\phi_4'
\\
\x^2:\ 0=\b_2(30\phi_1 -12\o\phi_2 -6\phi_3\phi_4)&
\!{}+\b_3(36\phi_0\phantom{\phi_4}-\phantom{1}9\phi_3\phi_3)&
\!{}+12t\phi_2 +6\phi_1' -6\o\phi_2' -3\phi_3\phi_4'
\\
\x^3:\ 0=\b_2(24\phi_2 -\phantom{1}9\o\phi_3 -8\phi_4\phi_4)&
\!{}+\b_3(30\phi_1\phantom{\phi_4} -15\phi_3\phi_4) &
\!{}+\phantom{1}9t\phi_3+6\phi_2' -6\o\phi_3' -4\phi_4\phi_4'
\\
\x^4:\ 0=\b_2(18\phi_3 -\phantom{1}6\o\phi_4)& 
\!{}+\b_3(24\phi_2\phantom{\phi_4} -\phantom{1}6\phi_4\phi_4) &
\!{}+\phantom{1}6t\phi_4 +6\phi_3' -6\o\phi_4'.
\end{array}
\ee
We see from this list of equations that the only way to solve for 
the remaining two $\b_a$'s is to make conditions on relations among 
the $\phi_k$'s.  By eliminating $\b_2$ using the $\x^4$ equation
and then $\b_3$ using the $\x^3$ equation, we get constraints which
are of the lowest order in $\o$.  Thus, if both
\bea\label{assumptions}
0 &\not\equiv& 3\phi_3-\o\phi_4, \\
0 &\not\equiv& 30\phi_1(3\phi_3-\o\phi_4) -96\phi_2^2
+56\phi_2\phi_4^2+36\o\phi_2\phi_3-45\phi_3^2\phi_4 
+6\o\phi_3\phi_4^2 -8\phi_4^4 ,\nonumber
\eea
then there are three remaining relations among the $\phi_k(\o)$'s:
\bea\label{eqnlist4}
0&=&
576 t [0 2 2]-540 t [0 1 3]-27 t [1 3 3 3]
+48 t [1 2 3 4]+324 t [0 3 3 4]-20 t [1 1 4 4]-384 t [0 2 4 4]
\nonumber\\&&
-4 t [1 3 4 4 4]+64 t [0 4 4 4 4]-192 w [2 2 0']
+180 w [1 3 0']+72 w^2 [2 3 0']-60 w^2 [1 4 0']
\nonumber\\&&
-90 w [3 3 4 0']+112 w [2 4 4 0']+12 w^2 [3 4 4 0']
-16 w [4 4 4 4 0']+144 w [0 2 2']-18 [1 3 3 2']
\nonumber\\&&
+16 [1 2 4 2']+36 [0 3 4 2']+6 w [1 3 4 2']-48 w [0 4 4 2']
-4 [1 4 4 4 2']-180 w [0 1 3']-144 w^2 [0 2 3']
\nonumber\\&&
+24 [1 2 3 3']+9 w [1 3 3 3']-20 [1 1 4 3']-48 [0 2 4 3']
-16 w [1 2 4 3']+72 w [0 3 4 3']-6 w^2 [1 3 4 3']
\nonumber\\&&
+48 w^2 [0 4 4 3']+2 [1 3 4 4 3']+16 [0 4 4 4 3']
+4 w [1 4 4 4 3']+180 w^2 [0 1 4']-32 [1 2 2 4']
\nonumber\\&&
+30 [1 1 3 4']-12 w [1 2 3 4']+9 w^2 [1 3 3 4']
+10 w [1 1 4 4']-48 w [0 2 4 4']-108 w^2 [0 3 4 4']
\nonumber\\&&
-3 [1 3 3 4 4']+8 [1 2 4 4 4']-24 [0 3 4 4 4']
-4 w [1 3 4 4 4']+16 w [0 4 4 4 4'],
\nonumber\\
0&=&
240 t [1 2 2]-225 t [1 1 3]-216 t [0 2 3]-27 t [2 3 3 3]
+180 t [0 1 4]+48 t [2 2 3 4]+126 t [1 3 3 4]
\nonumber\\&&
-172 t [1 2 4 4]-36 t [0 3 4 4]-4 t [2 3 4 4 4]
+24 t [1 4 4 4 4]+96 [2 2 0']-90 [1 3 0']-36 w [2 3 0']
\nonumber\\&&
+30 w [1 4 0']+45 [3 3 4 0']-56 [2 4 4 0']-6 w [3 4 4 0']
+8 [4 4 4 4 0']-96 w [2 2 1']+90 w [1 3 1']
\nonumber\\&&
+36 w^2 [2 3 1']-30 w^2 [1 4 1']-45 w [3 3 4 1']
+56 w [2 4 4 1']+6 w^2 [3 4 4 1']-8 w [4 4 4 4 1']
\nonumber\\&&
-144 [0 2 2']+60 w [1 2 2']-18 [2 3 3 2']+16 [2 2 4 2']
+9 [1 3 4 2']+6 w [2 3 4 2']+36 [0 4 4 2']
\nonumber\\&&
-18 w [1 4 4 2']-4 [2 4 4 4 2']+180 [0 1 3']
-75 w [1 1 3']+144 w [0 2 3']-60 w^2 [1 2 3']+24 [2 2 3 3']
\nonumber\\&&
+9 w [2 3 3 3']-32 [1 2 4 3']-16 w [2 2 4 3']-90 [0 3 4 3']
+33 w [1 3 4 3']-6 w^2 [2 3 4 3']-36 w [0 4 4 3']
\nonumber\\&&
+18 w^2 [1 4 4 3']+2 [2 3 4 4 3']+4 [1 4 4 4 3']
+4 w [2 4 4 4 3']-180 w [0 1 4']+75 w^2 [1 1 4']-32 [2 2 2 4']
\nonumber\\&&
+30 [1 2 3 4']-12 w [2 2 3 4']+9 w^2 [2 3 3 4']
+96 [0 2 4 4']-18 w [1 2 4 4']+90 w [0 3 4 4']
\nonumber\\&&
-42 w^2 [1 3 4 4']-3 [2 3 3 4 4']+8 [2 2 4 4 4']
-6 [1 3 4 4 4']-4 w [2 3 4 4 4']-240 [4 4 4 4' ] +8 w [1 4 4 4 4'] ,
\nonumber\\
0&=&
384 t [2 2 2]-720 t [1 2 3]+324 t [0 3 3]-81 t [3 3 3 3]
+300 t [1 1 4]-288 t [0 2 4]+324 t [2 3 3 4]
\nonumber\\&&
-224 t [2 2 4 4]-120 t [1 3 4 4]+96 t [0 4 4 4]-12 t [3 3 4 4 4]
+32 t [2 4 4 4 4]+192 [2 2 1']-180 [1 3 1']
\nonumber\\&&
-72 w [2 3 1']+60 w [1 4 1']+90 [3 3 4 1']-112 [2 4 4 1']
-12 w [3 4 4 1']+16 [4 4 4 4 1']-240 [1 2 2']
\nonumber\\&&
-96 w [2 2 2']+216 [0 3 2']+180 w [1 3 2']+72 w^2 [2 3 2']
-54 [3 3 3 2']-72 w [0 4 2']-60 w^2 [1 4 2']
\nonumber\\&&
+48 [2 3 4 2']-72 w [3 3 4 2']+60 [1 4 4 2']
+88 w [2 4 4 2']+12 w^2 [3 4 4 2']-12 [3 4 4 4 2']
\nonumber\\&&
-16 w [4 4 4 4 2']+300 [1 1 3']-288 [0 2 3']
+120 w [1 2 3']-96 w^2 [2 2 3']-108 w [0 3 3']
\nonumber\\&&
+72 [2 3 3 3']+27 w [3 3 3 3']+72 w^2 [0 4 3']
-210 [1 3 4 3' ] +12 w [2 3 4 3']-18 w^2 [3 3 4 3']
\nonumber\\&&
+96 [0 4 4 3']-60 w [1 4 4 3']+24 w^2 [2 4 4 3']
+6 [3 3 4 4 3']+12 w [3 4 4 4 3']-300 w [1 1 4']
\nonumber\\&&
+288 w [0 2 4']+120 w^2 [1 2 4']-108 w^2 [0 3 4']
-96 [2 2 3 4']+90 [1 3 3 4']-36 w [2 3 3 4']
\nonumber\\&&
+27 w^2 [3 3 3 4']+160 [1 2 4 4']-64 w [2 2 4 4']
-144 [0 3 4 4']+180 w [1 3 4 4']-60 w^2 [2 3 4 4']
\nonumber\\&&
-9 [3 3 3 4 4' ] -48 w [0 4 4 4']+24 [2 3 4 4 4']
-12 w [3 3 4 4 4']-40 [1 4 4 4 4']+16 w [2 4 4 4 4'] ,
\eea
where we have introduced the shorthand notations
\be
[ij\cdots k] := \phi_i \phi_j \cdots \phi_k, \qquad 
[ij\cdots k\ell'] := \phi_i \phi_j \cdots \phi_k \phi_\ell' .
\ee
Now, since the $\phi_k(\o)$ are polynomials in the $\o$ of orders
determined from (\ref{fexp}),
\bea\label{phiexpA}
\phi_4(\o)&=&\varphi_{40}+\varphi_{41}\o,
\nonumber\\
\phi_3(\o)&=&\varphi_{30}+\varphi_{31}\o+\varphi_{32}\o^2,
\nonumber\\
\phi_2(\o)&=&\varphi_{20}+\varphi_{21}\o+\varphi_{22}\o^2+\varphi_{23}\o^3,
\nonumber\\
\phi_1(\o)&=&\varphi_{10}+\varphi_{11}\o+\varphi_{12}\o^2 
+\varphi_{13}\o^3+\varphi_{14}\o^4,
\nonumber\\
\phi_0(\o)&=&\varphi_{00}+\varphi_{01}\o+\varphi_{02}\o^2 
+\varphi_{03}\o^3+\varphi_{04}\o^4+\varphi_{05}\o^5,
\eea
upon substitution in (\ref{eqnlist4}) gives three polynomials
in $\o$ of degree 9, 10, and 11, respectively.  Equating the
coefficients of each power of $\o$ to zero then gives 33 polynomial
equations in the 20 unknown coefficients, $\varphi_{k\ell}$, of the 
curve, as well as the dimension, $[v]$, through $t$.  It is
not too hard to see that these equations are graded homogenous in
the $\varphi_{k\ell}$ if $\varphi_{k\ell}$ is given weight
$6-k-\ell$.  This is a reflection of the inhomogeneous scaling
symmetry (\ref{Bgauge}), and means that any solution will actually
be at least a one (complex) parameter set of solutions.

Before we discuss methods for solving such systems of equations, we 
comment on the role played by the constraints (\ref{assumptions})
that were assumed to eliminate the $\b_a(\o)$'s.  First, these
constraints are fairly weak, since they demand that the right
sides, which are polynomials in $\o$ do not vanish identically.
Equivalently, they demand that not all the coefficients of the
various powers of $\o$ (which are themselves polynomials in the 
$\varphi_{k\ell}$) vanish.  Nevertheless, any solutions for
which these constraint are not met must be rejected.  Furthermore,
we must do a separate analysis and derivation of the $\varphi_{k\ell}$
equations in the case that one or both of the constraints is
not satisfied.  For example, if the first constraint in (\ref{assumptions})
were not satisfied, that is, if
\be\label{assump1}
0=3\phi_3-\o\phi_4,
\ee
then the $\x^4$ equation (\ref{eqnlist2}) cannot be solved for $\b_2$. 
Instead, we must solve the $\x^3$ equation for $\b_2$, giving the
new constraint
\be\label{assump2}
0 \not\equiv 24\phi_2 -9\o\phi_3 -8\phi_4\phi_4.
\ee
Then $\b_3$ can also be eliminated, giving another constraint,
and leaving three relations among the $\phi_k(\o)$'s different
from (\ref{eqnlist4}).  Expanding these and (\ref{assump1}) then
gives a new set of polynomial equations for the $\varphi_{k\ell}$
and $t$.  Clearly, this whole process must again be repeated
by assuming each new constraint is violated, giving a (finite)
tree of sets of equations that must be solved.

There exists an algorithm for solving simultaneous sets of polynomial
equations, known as Buchberger's algorithm for computing Groebner bases 
of polynomial ideals \cite{clo98}.  This can be used, in principle, to 
solve our system by giving an equivalent set of polynomials in which 
each variable ($t$, $\varphi_{k\ell}$) can be eliminated sequentially
by solving for the roots of a polynomial in a single variable.
Unfortunately, Buchberger's algorithm generates a large number of 
very high order polynomials with very very large coefficients as 
intermediate steps, and, for the set of equations derived from 
(\ref{eqnlist4}), is not workable on our PCs using implementations 
such as {\it Mathematica} or {\it Macaulay2} \cite{mac2}.  As alternatives 
to using a dedicated computer (or to being more dedicated computer
users), some ideas that may simplify this problem are mentioned in 
section \ref{s5}.  Also, some simplifying guesses which lead to a 
substantial, but still presumably incomplete, list of solutions to
our equations are described in appendix B.

\section{Known solutions}

A long list of solutions to the set of polynomial equations described in
appendix A can be found by making some simplifying guesses for the
form of the $\b_a$.  Because guesses are made, there is no
assurance that we have found all the solutions in this way.

It turns out that the set of equations becomes manageable if we assume
that two of the $\b_a$'s vanish.  A lengthy case-by-case analysis
of the equations then gives the list of solutions reproduced in table
\ref{tab.crv}.  Before explaining the notation
used in those tables, let's first remark on the $\b_{a_1}=\b_{a_2}=0$
{\it anzatz}.  Of the six possible choices of $(a_1,a_2)$, one pair,
$(0,3)$, gives all the solutions appearing in table \ref{tab.crv}.  
It is not hard to show that the assumption that $\b_0=\b_3=0$ is 
equivalent to the assumption that the $\b_a$ have the simple form 
$\b_a = \ell_a \o^{1-a}$, where the $\ell_a$ are constants.  Of the 
other choices, only $(a_1,a_2)=(2,3)$ gives new solutions: three 
curves with $[v]=-3/2$, $-5/3$, and $-2$; and these are all unphysical 
since they have negative scaling dimensions.

\TABLE[ht]
{\begin{tabular}{||ll|l|l|l||} \hline
\multicolumn{3}{||c|}{Dimensions}& 
\multicolumn{2}{c||}{Curve\ \ $\eta^2=\a(\o)\,\phi(\o,\x)$}\\
$[v]$&	& $[v]/[u]\in$			&$\phi=\x^6+\ldots$
&$\a=$		\\ \hline\hline
1$^*$&	& $\Z>1$					&$\t_1\x^4+\t_2\x^3+\t_3\x^2+\t_4\x+1$
&1			\\ \hline
6/5	&	& $\Z/5\ge6/5$			&$\o$
&$\o^{-1/3}$	\\ \hline
5/4	&	& $\Z/4\ge5/4$			&$\o\x$	
&$\o^{-2/5}$	\\ \hline
3/2 &$a$	& $\Z/5\ge6/5$			&$6\x-5\o$
&1			\\ 
	&$b$	& $\Z/2\ge3/2$			&$\t\o\x^3+\o^2$
&$\o^{-2/3}$	\\ 
	&$b'$& $\Z\pm{1\over4}>3/2$	&$\o^2$
&$\o^{-2/3}$	\\ \hline
5/3	&$a$	& $\Z/4\ge5/4$			&$\x(5\x-4\o)$
&1			\\ 
	&$b$	& $\Z/3\ge5/3$			&$\o^2\x$
&$\o^{-4/5}$	\\ \hline
15/8&	& $\Z/4\ge3/2$			&$\o(6\x-5\o)$
&$\o^{-2/5}$	\\ \hline
2	&	& $\Z\ge2$				&$\t_1\o\x^4+\t_2\o^2\x^2+\o^3$
&$\o^{-1}$	\\
	&$'$	& $\Z\pm{1\over3}\ge2$	&$\o^3$
&$\o^{-1}$	\\ \hline
20/9&	& $\Z/3\ge5/3$			&$\o\x(5\x-4\o)$
&$\o^{-1/2}$	\\ \hline
5/2	&$a$	& $\Z/3\ge2$				&$\o^2(6\x-5\o)$
&$\o^{-4/5}$	\\ 
	&$b$	& $\Z/2\ge5/2$			&$\o^3\x$
&$\o^{-6/5}$	\\ \hline
3	&$a$	& $\Z/2\ge3/2$			&$\t\x^3(3\x-2\o)+(3\x-2\o)^2$
&1			\\ 
	&$a'$& $\Z\pm{1\over4}\ge3/2$	&$(3\x-2\o)^2$
&1			\\
	&$b$	& $\Z\ge3$				&$\t\o^2\x^3+\o^4$
&$\o^{-4/3}$	\\ 
	&$b'$& $\Z\pm{1\over2}\ge3$	&$\o^4$
&$\o^{-4/3}$	\\ \hline
10/3	&	& $\Z/2\ge5/2$			&$\o^2\x(5\x-4\o)$
&$\o^{-1}$	\\ \hline
15/4	&	& $\Z/2\ge3$				&$\o^3(6\x-5\o)$
&$\o^{-6/5}$	\\ \hline
4	&	& $\Z/3\ge2$				&$\o(3\x-2\o)^2$
&$\o^{-1/2}$	\\ \hline
5	&$a$	& $\Z/3\ge5/3$			&$\x(5\x-3\o)^2$
&1			\\
	&$b$	& $\Z\ge5$				&$\o^4\x$
&$\o^{-8/5}$	\\ \hline
6	&$a$	&$\Z\ge3$				&$\t\o\x^3(3\x-2\o)+\o^2(3\x-2\o)^2$
&$\o^{-1}$	\\ 
	&$a'$& $\Z\pm{1\over2}\ge3$	&$\o^2(3\x-2\o)^2$
&$\o^{-1}$	\\
	&$b$	& $\Z\ge6$				&$\o^5$
&$\o^{-5/3}$	\\ \hline
20/3	&	& $\Z\ge5$				&$\o^3\x(5\x-4\o)$
&$\o^{-3/2}$	\\ \hline
15/2&$a$	& $\Z/2\ge5/2$			&$\o\x(5\x-3\o)^2$
&$\o^{-2/3}$	\\ 
	&$b$	& $\Z\ge6$				&$\o^4(6\x-5\o)$
&$\o^{-8/5}$	\\ \hline
12	&	& $\Z\ge6$				&$\o^3(3\x-2\o)^2$
&$\o^{-3/2}$	\\ \hline
15	&	& $\Z\ge5$				&$\o^2\x(5\x-3\o)^2$
&$\o^{-4/3}$	\\ \hline
\end{tabular}
\caption{Potentially physical solutions of the integrability equation.
($^*$For $[v]=1$, we can have $r\in\Z/n>1$ for $n=2,3,4$, or 
$5$ when some $\t_i=0$.)\label{tab.crv}}}

In table \ref{tab.crv} we list our solutions.  We give a list of
the curves written in the form
\be
\eta^2 = \a(\o) \left[ \x^6 + \phi(\x,\o) \right]
\ee
in terms of the scaling variables whose definitions (\ref{sca}) we 
recall here:
\be\label{scaB}
\o := -r v u^{-r} ,\qquad
\x := u^{1-r} x ,\qquad
\eta := u^{1-2r+rs} y,\qquad
r := [v]/[u] ,\qquad s := 1/[v].
\ee
We have also used our freedom (\ref{Bgauge}) to rescale one 
coefficient in $\phi(\x,\o)$; any other undetermined coefficients, 
$\t_i$, are then dimensionless parameters in the curve (marginal 
deformations of the corresponding $N=2$ theory).

The curve prefactor $\a(\o)$ is determined up to an overall
coefficient by the equations determining the curve.  The overall
factor can be scaled to 1 using the $A$-rescaling freedom in
(\ref{holrep}).  It turns out that for all our curves, the exponent 
appearing in $\a$ is simply related to the form of the curve.  If 
the term in the curve with the highest power of $\x$ is proportional 
to $\o^j\x^k$, then the prefactor will be $\alpha=\o^{-p}$ with $p 
= 2j/(6-k)$.

For each curve we give the scaling dimension $[v]$, 
which is found from the equations determining the curve. 
The first column lists the scaling dimension, $[v]$, and the
third and fourth columns give the 
curve; these are all data determined solely by the integrability
equation.  Note that some of the curves can depend on free parameters,
$\t_i$.  We will refer to
the different curves in what follows by their value of $[v]$.  In the
cases where different curves have the same $[v]$, we append a letter
($a, b, \ldots$) to differentiate them.  When a curve is a specialization
of another one (\eg\ by setting some parameters $\t_i=0$), we denote
it with a prime.  Thus, for example, the second curve in table \ref{tab.crv} 
with $[v]=3$ will be referred to as the $3a'$ curve in what follows.

The second column lists the allowed values of the $[u]$ scaling
dimension by listing the allowed values of the ratio $r$.  $r$ is
primarily constrained by the monodromy conditions on the
curve, which implied that the only allowed $u$ and $v$
dependence was of the form $y^2=a(u,v)[x^6+\sum_{k=0}^5 
f_k(u,v) x^k]$ with the $f_k$ satisfying (\ref{fexp}), in which 
all the powers of $u$ and $v$ must be non-negative integers.
Also, as discussed after (\ref{fexp}), this together with the
scaling assumption implies that $r>1$.
These restrictions, together with the form of the curve given in the
table, then determine the allowed set of $r$ values.  Let us illustrate
this on an example.  Consider the $3a$ curve in table \ref{tab.crv}.
It has terms proportional to $\x^4$, $\o\x^3$, $\x^2$, $\o\x$,
and $\o^2$.  Since by (\ref{scaB}) $\x\propto x$ and $\o \propto v$, 
these terms will be proportional to, respectively, $x^4$, $vx^3$, $x^2$,
$vx$, and $v^2$, times some factors of $u$ determined by dimensional
scaling.  Indeed, the powers of $u$ can be read off from (\ref{fexp})
to be $u^{2r-2}x^4$, $vu^{2r-3}x^3$, $u^{4r-4}x^2$, $vu^{4r-5}x$, 
and $v^2u^{4r-6}$.  Then the condition that the powers of $u$ be
non-negative integers, restricts $r\in\Z/2>1$, $r\in\Z/2\ge3/2$,
$r\in\Z/4>1$, $r\in\Z/4\ge5/4$, and $r\in\Z/4\ge3/2$, respectively
for each term.  Since all these terms appear in the curve, the
allowed values of $r$ are the intersection of these sets, which in
this case is $r\in\Z/2\ge3/2$, as listed in the table.  Note that
if the first two terms are set to zero (by setting the parameter 
$\t=0$ in the curve), then the constraints on $r$ are loosened to allow
the extra values $r\in\{(\Z/4)\backslash (\Z/2)\}=\{\Z\pm{1\over4}\}
\ge3/2$; these are listed separately as the $3a'$ curve in 
the table.

Table \ref{tab.crv} does not list all the solutions we found of
the BPS integrability equations, though.  In addition we found
a smaller set of manifestly unphysical solutions.  These are 
solutions which, though solutions to the BPS integrability 
equations, violate some other property required of scale-invariant 
RSK geometries: either they do not resolve the singularity on the 
Coulomb branch, or they violate our scaling assumptions.
Any curves that do not resolve the singularity on the Coulomb 
branch---ones which remain singular for all $u$ and $v$---must 
be thrown out simply because Seiberg-Witten curves only have a 
well-defined physical interpretation describing an $N=2$ $\U(1)^2$ 
low energy effective action when they are non-singular. 
We found six such singular solutions with $[v]= 4/3$, $2$, $3$,
$4$, and $6$, and one for which $[v]$ can be arbitrary.  
Our demand for solutions which scale to a singularity 
at $u=v=0$ requires, in particular, that $\Re [v] >0$.  The solutions 
we found with $[v]=-10$, $-6$, $-5$, $-3$, $-2$, $-5/3$, and $-3/2$
thus simply do not describe a singularity at the origin of $\MM$, and 
so are discarded.  Similarly, two solutions with $[v]=\infty$ also have 
no scaling interpretation, and so are unphysical.

\section*{Acknowledgments}
It is a pleasure to thank A. Buchel, R. Gass, R. Karp, A. Klemm, M. Kleban, 
D. Morrison, R. Rabad\'an and E. Witten for helpful comments and discussions.  
PCA and JW are supported in part by DOE grant DOE-FG02-84ER-40153.  
PCA is also supported by an IBM Einstein Endowed Fellowship.  
MC is supported in part by Kentucky NSF EPSCoR award EPS-9874764.Ê 
ADS is supported by DOE grant DEÐFG01Ð00ER-45832 and NSF grant PHYÐ0245214.

\end{document}